\documentclass[reqno,centertags,11pt]{amsart}
\usepackage{amsmath,amsthm,amscd,amssymb}
\usepackage{latexsym}
\usepackage{pstricks}
\usepackage{pst-node}
\usepackage {verbatim}

\newcommand{\N}{{\mathbb{N}}}

\newcommand{\R}{{\mathbb{R}}}

\newcommand{\C}{{\mathbb{C}}}

\newcommand{\Z}{{\mathbb{Z}}}

\newcommand{\f}{\frac}

\newcommand{\ol}{\overline}

\newcommand{\wti}{\widetilde  }

\newcommand{\ran}{\text{\rm{ran}}}

\newcommand{\hatt}{\widehat}
\newcommand{\beq}{\begin{equation}}
\newcommand{\eeq}{\end{equation}}
\newcommand{\bdm}{\begin{displaymath}}
\newcommand{\edm}{\end{displaymath}}
\newcommand{\ba}{\begin{align}}
\newcommand{\ea}{\end{align}}
\newcommand{\bpf}{\begin{proof}}
\newcommand{\epf}{\end{proof}}

\newcommand{\la}{\langle}
\newcommand{\ra}{\rangle}

\newcommand{\supp}{\mathrm{supp}\, }               
\newcommand{\dist}{\mathrm{dist}}               

\newcommand{\veps}{\varepsilon}

\newcommand{\re}{\mathrm{Re}}
\newcommand{\im}{\mathrm{Im}}

\newcommand{\id}{\mathbf{1}}                

\allowdisplaybreaks

\newcommand{\calG}{\mathcal{G}}

\newcommand{\calQ}{\mathcal{Q}}

\newtheorem{thm}{Theorem}
\newtheorem{prop}[thm]{Proposition}
\newtheorem{lem}[thm]{Lemma}

\theoremstyle{definition}

\newtheorem{remark}[thm]{Remark}
\newtheorem{remarks}[thm]{Remarks}

\newcounter{theoremi}[thm]

\newcommand{\itemthm}{\refstepcounter{theoremi} {\rm(\roman{theoremi})}{~}}

\numberwithin{thm}{section}
\numberwithin{equation}{section}

\newcounter{smalllist}

\newcounter{smallenum}
\newenvironment{SE}{\begin{list}{{\rm\arabic{smallenum})}}{%
\setlength{\topsep}{0mm}\setlength{\parsep}{0mm}\setlength{\itemsep}{0mm}%
\setlength{\labelwidth}{2em}\setlength{\leftmargin}{2em}\usecounter{smallenum}%
}}{\end{list}}
\begin{document}

\title[Non-local variational problems]{%
    On non-local variational problems with lack of compactness related to non-linear optics}
\author[D.~Hundertmark and Y.-R.~Lee]{Dirk Hundertmark and Young-Ran~Lee}
\address{Department of Mathematics, Altgeld Hall,
    and Institute for Condensed Matter Theory at the
    University of Illinois at Urbana--Champaign,
    1409 W.~Green Street, Urbana, IL 61801.}%
\email{dirk@math.uiuc.edu}%
\address{Department of Mathematics, Sogang University, Shinsu-dong 1,
    Mapo-gu, Seoul, 121-742, South Korea.}%
\email{younglee@sogang.ac.kr}

\thanks{\copyright 2009 by the authors. Faithful reproduction of this article,
        in its entirety, by any means is permitted for non-commercial purposes}
\thanks{Supported in part by NSF-grant DMS-0803120 (D.H.) and the National
Research Foundation of Korea (NRF)-grant 2009-0064945 (Y.-R.L.).}

\begin{abstract}
We give a simple proof of existence of solutions of the dispersion
management and diffraction management equations for zero average dispersion,
respectively diffraction. These solutions are found as maximizers of
non-linear and non-local variational problems which are invariant under
a large non-compact group. Our proof of existence of maximizer is
rather direct and avoids the use of Lions' concentration compactness
argument or Ekeland's variational principle.
\end{abstract}

\maketitle

\section{Introduction}\label{introduction}
\subsection{The variational problems}\label{sec:introduction1}
In this paper we are concerned with the existence of maximizers for two non-local non-linear variational problems,
    \beq\label{eq:max-cont}
        P_\lambda^\mathrm{c}:= \sup\Big(\calQ_\mu^\mathrm{c}(f,f,f,f) \big\vert\, f\in L^2(\R), \|f\|_{L^2}^2=\lambda \Big) ,
    \eeq
respectively,
    \beq\label{eq:max-discr}
        P_\lambda^\mathrm{d}:= \sup\Big(\calQ_\mu^\mathrm{d}(f,f,f,f) \big\vert\, f\in l^2(\Z), \|f\|_{l^2}^2=\lambda \Big) ,
    \eeq
for $\lambda>0$. Here the four-linear functionals $\calQ_\mu^{\mathrm{c}/\mathrm{d}}$ are given by
    \beq\label{eq:calQ-cont}
        \calQ_\mu^\mathrm{c}(f_1,f_2,f_3,f_4)
        :=
        \int_\R\int_\R \ol{(T_rf_1)(x)} (T_rf_2)(x)\ol{(T_rf_3)(x)} (T_rf_4)(x)\, dx \mu(dr)
    \eeq
in the continuous case, where $\mu$ is a suitable measure on $\R$, the operator $T_r:= e^{ir\partial_x^2}$ is the unitary solution operator for the free Schr\"odinger equation in one dimension, and $f_j\in L^2(\R)$ for
$j=1,2,3,4$,  respectively
    \beq\label{eq:calQ-discr}
        \calQ_\mu^\mathrm{d}(f_1,f_2,f_3,f_4)
        :=
        \sum_{x\in\Z}\int_\R \ol{(S_rf_1)(x)} (S_rf_2)(x)\ol{(S_rf_3)(x)} (S_rf_4)(x)\, \mu(dr)
    \eeq
in the discrete case, where, with $\Delta$ the discrete Laplacian given by
$\Delta f(x)= f(x+1)+f(x-1)-2f(x)$ we denote by $S_r:= e^{ir\Delta}$ the solution operator of the
free discrete Schr\"odinger equation in one dimension and $f_j\in l^2(\Z)$, $j=1,2,3,4$.

For the existence of maximizers of \eqref{eq:max-discr} we only need that $\mu$ is a bounded measure with bounded support, see Theorem \ref{thm:existence-discr} and for the existence of maximizers of \eqref{eq:max-cont} we need that the measure $\mu$ has a density $\psi$ lying in suitable $L^p$-spaces, see Theorem \ref{thm:existence-cont}.

Our interest in these variational problems stems from the fact that the maximizers of \eqref{eq:max-cont}, respectively \eqref{eq:max-discr}, yield (quasi-)periodic breather type solutions, the dispersion management solitons, of the dispersion managed non-linear Schr\"odinger equation, respectively the diffraction management solitons for the diffraction managed discrete non-linear Schr\"odinger equation. The dispersion management solitons have attracted a lot of interest in the development of ultra--fast long--haul optical data transmission fibers and the diffraction management solitons where studied in some new discrete waveguide
array designs. We address the connection of the two variational problems above with non-linear optics
later in section \ref{sec:introduction2}.

The standard approach to show the existence of a maximizer of \eqref{eq:max-cont}, respectively, \eqref{eq:max-discr}, is to identify it as the strong limit of a suitable maximizing sequence, i.e.,
in the continuous case a sequence $(f_n)_{n\in\N}\subset L^2(\R)$ with $\|f_n\|_2^2=\lambda$ and
$ P_\lambda^\mathrm{c} = \lim_{n\to\infty} \calQ_\mu^\mathrm{c}(f_n,f_n,f_n,f_n)$.
The problem is that the two variational problems above are invariant under translations of $L^2(\R)$, respectively $l^2(\Z)$. The invariance of
$(l^2(\Z))^4\ni (f_1,f_2,f_3,f_4)\mapsto \calQ_\mu^\mathrm{d}(f_1,f_2,f_3,f_4)$ under simultaneous
translations of the $f_j$ follows simply from the invariance of the discrete Laplacian $\Delta$ and hence of
$S_r=e^{ir\Delta}$ under shifts.
Even worse, due to the Galilei invariance of the free  Schr\"odinger evolution the maximization problem \eqref{eq:max-cont} is invariant under translations \emph{and} boosts, i.e., translation in
momentum space, of $L^2(\R)$ for every sensible choice of the measure $\mu$, see the discussion in Appendix \ref{sec:Galilei}.
Thus the two variational problems above are invariant under a large non-compact group
of transformations leading to a \emph{loss of compactness}: maximizing sequences can easily
converge weakly to zero.

Under the assumption that the measure $\mu$ has density $\id_{[0,1]}$, that is, the uniform distribution on $[0,1]$, this loss of compactness in
the variational problem \eqref{eq:max-cont} was overcome by Kunze in \cite{Kunze04}, who used a very tricky application of the Lions' concentration compactness principle, \cite{Lions84}, first in Fourier-space and then in real space, to compensate for the loss of compactness due to shifts and boosts.
The existence of maximizers of \eqref{eq:max-discr}, again under the same assumption that
$\mu$ has density $\id_{[0,1]}$, was shown by Stanislavova in \cite{Stanislavova07}, using Ekeland's
variational principle \cite{Ekeland72,Ekeland73,Jabri}.

In this paper we give an alternative approach very much different from \cite{Kunze04}
and \cite{Stanislavova07} for the existence of maximizers for the variational problems
\eqref{eq:max-cont} and \eqref{eq:max-discr} which we believe is not only very natural but has
several additional advantages:
\begin{SE}
\item We show existence of maximizers of \eqref{eq:max-cont} under \emph{very weak} conditions
on the measure $\mu$, see Theorem \ref{thm:existence-cont}, and the existence of maximizers of
\eqref{eq:max-discr} under the \emph{weakest possible} assumption on $\mu$, see Theorem \ref{thm:existence-discr} and Remark \ref{rem:conditions}.
\item Our approach avoids the use of `heavy machinery' like Lions' concentration compactness
or Ekeland's variational principle from the calculus of variations and gives easily
more information about maximizing sequences. Whereas \cite{Kunze04} and \cite{Stanislavova07}
show that there exists at least one suitable maximizing sequence which is strongly converging, we show
that the loss of compactness is much milder than one would naively expect:
\emph{Any} maximizing sequence for \eqref{eq:max-cont}, respectively \eqref{eq:max-discr}, is tight,
i.e., stays in a compact subset of $L^2(\R)$, respectively $l^2(\Z)$, modulo translations and
boosts in $L^2$, respectively modulo translations in $l^2$, see
Propositions~\ref{prop:max-seq-tight-cont} and \ref{prop:max-seq-tight-discr}.
\item To conclude that a maximizer of \eqref{eq:max-cont}, respectively \eqref{eq:max-discr},
exists, we use a simple characterization of strong convergence in $L^2$, respectively
$l^2$, in terms of `weak convergence' and `tightness'.  This is done in Lemma
\ref{lem:strong-convergence-L2} and \ref{lem:strong-convergence-lp} whose proof is
rather straightforward and uses only some simple properties of compact operators.
\item We believe that these ideas will be useful in the study of other variational problems on $L^2$, respectively $l^2$.
\end{SE}

Our main results concerning the variational problems \eqref{eq:max-cont} and \eqref{eq:max-discr}
are

\begin{thm}[Existence, continuous case]\label{thm:existence-cont}
Assume that the measure $\mu$ has a density $\psi$ with $\psi\in L^2(\R)$. Then the variational problem \eqref{eq:max-cont} is well-posed, i.e., $P_\lambda^\mathrm{c}<\infty$ for all $\lambda>0$.
Moreover, if the density $\psi \in L^2(\R)\cap L^4(\R)\cap L^4(\R, t^2dt)$,
then for any $\lambda>0$, there exists a maximizer for the variational problem \eqref{eq:max-cont},
i.e., there exists $f\in L^2(\R)$, $\|f\|_2^2=\lambda$, such that
    \bdm
       \calQ_\mu^\mathrm{c}(f,f,f,f) = \sup\Big(\calQ_\mu^\mathrm{c}(g,g,g,g) \big\vert\, g\in L^2(\R), \|g\|_2^2=\lambda\Big) .
    \edm
This maximizer is also a solution of the dispersion management equation
\eqref{eq:GT-weak-zero-cont} for some Lagrange multiplier $\omega>0$.
\end{thm}

In the discrete case we have an existence result under the `weakest possible' assumption
on the measure $\mu$.
\begin{thm}[Existence, discrete case]\label{thm:existence-discr}
The variational problem \eqref{eq:max-discr} is well posed if $\mu$ is a bounded measure.
If, in addition, the measure $\mu$ has bounded support, then for any $\lambda>0$, there exists
a maximizer for the variational problem \eqref{eq:max-discr}.
This maximizer is also a solution of the diffraction management equation \eqref{eq:GT-weak-zero-discr}
for some Lagrange multiplier $\omega>0$.
\end{thm}

\begin{remarks}
\itemthm\label{rem:conditions}
    As we will see in section \ref{sec:introduction2} the requirement that $\mu$ is a probability
    measure with bounded support arises naturally in the study of diffraction management solitons.
    In this sense, the condition on $\mu$ in Theorem \ref{thm:existence-discr} is optimal and the
    assumptions on the density of $\mu$ in Theorem \ref{thm:existence-cont} are not very restrictive since
    in this case $\psi\in L^1(\R)$ with compact support and thus, by interpolation, the assumptions in
    Theorem \ref{thm:existence-cont} reduce to the additional requirement that $\psi\in L^4(\R)$.
    In particular, the density $\psi$ can have some strong local singularities which, via
    \eqref{eq:psi-def} below, yields existence of breather type solutions in dispersion managed glass
    fiber cables under mild conditions on the dispersion profile $d_0$ of the fiber. For example, any
    locally continuous dispersion profile $d_0$ which is bounded away from zero is allowed, $d_0$
    can even have (isolated) zeros, as long as they are approached slowly enough. \\[0.2em]
\itemthm The existence Theorem \ref{thm:existence-cont} gives no further information about
    the regularity of the maximizer. If $\mu$ has density $\id_{[0,1]}$, Kunze's existence proof,
    \cite{Kunze04}, shows that the maximizer is bounded. In \cite{Stanislavova05} Stanislavova then
    showed that Kunze's maximizer is infinitely often differentiable.
    Only recently it was shown in \cite{HuLee2009} that any weak solution $f\in L^2(\R)$ of
    the dispersion management equation \eqref{eq:GT-weak-zero-cont} is a Schwartz function, i.e., it is infinitely often differentiable and all its derivatives decay faster than algebraically at infinity.
    All these result so far need that $\mu$ has density $\id_{[0,1]}$, but, as we shall see, the
    regularity result of \cite{HuLee2009} easily carries over to all $\mu$ considered in Theorem \ref{thm:existence-cont}, see Remark \ref{rem:regularity}.\\[0.2em]
\itemthm Again if $\mu$ has density $\id_{[0,1]}$, Lushnikov  gave convincing
    but non-rigorous arguments in \cite{Lushnikov04} that the maximizer of \eqref{eq:max-cont} should
    have the asymptotic form
    \bdm
        f(x)\sim A \cos(a_2x^2 + a_1 x + a_0) e^{-b |x|} \quad \text{ as } |x|\to\infty
    \edm
    for some $a_j$ and $b>0$. In particular, the maximizer should be exponentially decaying.
    In \cite{EHL2009} we show that any solution $f\in L^2(\R)$ of the dispersion management
    equation \eqref{eq:GT-weak-zero-cont}, so also any maximizer of \eqref{eq:max-cont},
    together with its Fourier transform is exponentially decaying if $\mu$ has density $\id_{[0,1]}$,
    confirming part of Lushinikov's conjecture.
    In particular, even though it is no longer an elliptic equation, the singular limit \eqref{eq:GT-weak-zero-cont} of the dispersion management equation still enjoys very strong
    regularity properties: any solution of it is analytic in a strip containing the real line
    under suitable conditions on the density of $\mu$. \\[0.2em]
\itemthm
    As already mentioned, the existence of maximizers of the discrete maximization problem
    \eqref{eq:max-discr} was shown in \cite{Stanislavova07} if $\mu$ has density $\id_{[0,1]}$.
    Moreover, \cite{Stanislavova07} shows that in this case the maximizer decays faster than
    algebraically at infinity. In \cite{HuLee-diffms} we show that under the conditions of the
    Existence Theorem \ref{thm:existence-discr} every maximizer and, more generally,
    any solution of the discrete Gabitov--Turitsyn equation \eqref{eq:GT-weak-zero-discr}
    for vanishing average diffraction, is even \emph{super-exponentially} decaying.
    More precisely, the bound
    \beq\label{eq:super-exp-decay}
        \limsup_{|x|\to\infty} \big((|x|+1)\ln(|x|+1)\big)^{-1} \ln|f(x)|\le -\frac{1}{4}
    \eeq
    holds for any solution $f\in l^2(\Z)$ of \eqref{eq:GT-weak-zero-discr} if $\mu$ is a bounded measure with compact support. Thus, unlike in the continuous case, we have the super-exponential decay  estimate
    \eqref{eq:super-exp-decay} for any solution under the same condition on the measure $\mu$ as needed for existence.
\end{remarks}

Our paper is organized as follows: In the next section we discuss how the maximization
problems \eqref{eq:max-cont}, respectively \eqref{eq:max-discr}, arise naturally in problems in non-linear optics where the dispersion, respectively diffraction, is strongly periodically
varied. In section \ref{sec:existence} we derive our main tools for the existence of maximizers
for the variational problems \eqref{eq:max-cont} and \eqref{eq:max-discr}.
In Proposition \ref{prop:max-seq-tight-cont} we show that any maximizing sequence
for \eqref{eq:max-cont} can be shifted and boosted, so that it is tight both in real
and Fourier space. A similar result, Proposition \ref{prop:max-seq-tight-discr}, is shown
in the discrete case. These two propositions are the key for avoiding the use of Lions'
concentration compactness argument or Ekeland's variational principle.
The core of the argument is given in Lemmata \ref{lem:fat-tail-bound-cont}, respectively, Lemma \ref{lem:fat-tail-bound-discr}, which show that a near maximizer cannot break up in real and Fourier space.
Strong convergence of suitably translated and boosted, respectively translated,
maximizing sequences then follows from a simple characterization of strong convergence
in $L^2$, respectively $l^2$, in Lemma \ref{lem:strong-convergence-L2} and \ref{lem:strong-convergence-lp}.
The proof of Lemma \ref{lem:fat-tail-bound-cont} is based on multi-linear refinement
of the Strichartz inequality in real and Fourier space. These refinements are extensions
of the multi-linear estimates developed in \cite{HuLee2009} and discussed in Appendix
\ref{sec:mult-linear-cont}. In the discrete case, we use the multi-linear estimates
developed in \cite{HuLee-diffms}, see Appendix \ref{sec:mult-linear-discr}.

\subsection{The connection with non-linear optics}\label{sec:introduction2}

Our main motivation for studying \eqref{eq:max-cont}, respectively \eqref{eq:max-discr},
comes from the fact that the maximizers of these two variational problems are related to
breather-type solutions of the dispersion managed non-linear Schr\"odinger equation
 \beq\label{eq:NLS-cont}
      i \partial_t u = -d(t) \partial_x^2 u - |u|^2 u,
 \eeq
respectively its discrete version,
 \beq\label{eq:NLS-discr}
    i \partial_t u(t,x) = - d(t) (\Delta u)(t,x) - |u(t,x)|^2 u(t,x), \quad x\in\Z  ,
 \eeq
where the dispersion/diffraction $d(t)$ is parametrically modulated. The continuous
non-linear Schr\"odinger equation \eqref{eq:NLS-cont}, respectively its discrete version
\eqref{eq:NLS-discr}, describe a wide range of different physical phenomena in such diverse
areas as solid states physics, some biological systems, Bose-Einstein condensation, and
continuous and discrete non-linear optics, e.g., glass-fiber cables and optical waveguide
arrays, see, e.g., \cite{CBG-J94,Davydov73,Scott85,SKES03,TS98}.

In non-linear optics \eqref{eq:NLS-cont} describes the evolution of a pulse in a frame moving
with the group velocity of the signal through a glass fiber cable, see \cite{SulemSulem}.
As a \emph{warning}: with our choice of notation the variable $t$ denotes the position along the glass fiber cable  and $x$ the (retarded) time. Hence $d(t)$ is \emph{not varying in time} but  denotes indeed a dispersion \emph{varying along} the optical cable.  The discrete version \eqref{eq:NLS-gen} describes an array of wave-guides where $t$ is the distance along the waveguide, the now discrete variable $x\in\Z$ denotes the location of an array element, and $d(t)$ the total diffraction along the waveguide. For physical reasons it is not a restriction to assume that $d$ is piecewise constant, but we will not make this assumption
in this paper.

In the continuous case the dispersion management idea, i.e., the possibility to periodically manage
the dispersion by put alternating sections with positive and negative dispersion together in an
optical glass-fiber cable to compensate for dispersion of the signal was predicted by Lin, Kogelnik,
and Cohen already in 1980, see \cite{LKC80}, and then implemented by Chraplyvy and Tkach for which they
received the Marconi prize in 2009. The periodically varying dispersion creates a new optical fiber type enabling the development of long--haul optical fiber transmission systems with record breaking capacities beyond one Terabit/second per fiber which equates to a 100-fold capacity increase in the last ten years,
\cite{AB98, CGTD93, Chraplyvy-etal95, GT96a, GT96b, KH97, Kurtzge93, LK98, LKC80, MM99, MGLWXK03, MMGNMG-NV99}.  Thus dispersion management technology has been of fundamental importance for ultra-high speed data transfer
through glass fiber cables over intercontinental distances and is now widely used commercially.
For a review see \cite{TDNMSF99, Tetal03}.
Discrete solitons in an optical waveguide array, on the other hand,
were theoretically predicted in \cite{CJ88}. Nearly a decade later they were experimentally studied, \cite{ESMBA98}, and as in the continuous case localized stable non-linear waves were found.
Recently a zigzag diffraction management geometry in discrete optical waveguides was
proposed in \cite{ESMA00} in order to create low power stable discrete pulses which can be more easily observed experimentally.

In both cases, the periodic modulation of the dispersion, respectively diffraction, can be described
by the ansatz
 \beq\label{eq:dispersion}
    d(t) = \veps^{-1} d_0(t/\veps) + d_{av} .
 \eeq
Here  $d_{av} \ge 0$ is the average component and $d_0$ its mean zero part which, by scaling,
we can assume to have period two.
For small $\veps$ the equation \eqref{eq:dispersion} describes a fast strongly varying dispersion, respectively diffraction, which corresponds to  the regime of \emph{strong} dispersion, respectively diffraction, management.

Since \eqref{eq:NLS-cont} and \eqref{eq:NLS-discr} are formally very similar we will combine them into the equation
    \beq\label{eq:NLS-gen}
        i \partial_t u = d(t) A u - |u|^2 u
    \eeq
on the Hilbert space $X$ where we call the choice $X=L^2(\R)$ and
$A=-\partial_x^2= -\tfrac{\partial^2}{\partial x^2}$ the continuous case
and the discrete case is given by  $X=l^2(\Z)$ and $A=-\Delta$ the discrete Laplacian.
We seek to rewrite \eqref{eq:NLS-gen} into a more amenable form in order to find breather
type solutions.
Let  $D(t)= \int_{-1}^t d_0(s)\, ds$ and note that as long as  $d_0$ is locally
integrable and has period two with mean zero, $D$ is also periodic with period two.
Furthermore,  $U_r= e^{-irA}$ is a unitary operator and thus the
unitary family $t\mapsto U_{D(t/\veps)}$ is periodic with period $2\veps$.
Making the ansatz $u(t,x)= (U_{D(t/\veps)}v(t,\cdot))(x)$ in \eqref{eq:NLS-gen}, a short calculation
shows
    \beq\label{eq:NLS-gen-2}
        i\partial_t v= d_{\text{av}}A v - U_{D(t/\veps)}^{-1}\big[|U_{D(t/\veps)} v|^2 U_{D(t/\veps)} v\big]
    \eeq
which is equivalent to \eqref{eq:NLS-gen} and still a non-autonomous equation.

For small $\veps$, that is, in the regime of strong dispersion/diffraction
management, $U_{D(t/\veps)}$ is fast oscillating in the variable $t$, hence the
solution $v$ should evolve on two widely separated time-scales, a slowly evolving part
$v_{\text{slow}}$ and a fast, oscillating part which is hopefully small.
Analogously to Kapitza's treatment of the unstable pendulum which is stabilized by fast
oscillations of the pivot, see \cite{LandauLifshitz}, the effective equation for the slow part
$v_{\text{slow}}$ was derived by Gabitov and Turitsyn \cite{GT96a,GT96b} in the continuous case
and in \cite{AM01,AM02,AM03} in the discrete case. It is given by integrating the fast
oscillating term containing $U_{D(t/\veps)}$ over one period in $t$,
    \beq
    \begin{split}\label{eq:GT-time-version-1}
        i\partial_t v_{\text{slow}}
        &=  d_{\text{av}}A v_{\text{slow}}
            - \frac{1}{2\veps}\int_{-\veps}^\veps
                U_{D(r/\veps)}^{-1}\big[|U_{D(r/\veps)} v_{\text{slow}}|^2 U_{D(r/\veps)} v_{\text{slow}}\big]
              \, dr \\
        &= d_{\text{av}}A v_{\text{slow}}
            - \frac{1}{2}\int_{-1}^1
                U_{D(r)}^{-1}\big[|U_{D(r)} v_{\text{slow}}|^2 U_{D(r)} v_{\text{slow}}\big]
              \, dr .
    \end{split}
    \eeq
This averaging procedure leading to \eqref{eq:GT-time-version-1} was rigorously justified
for the profile $d_0=\id_{[-1,0)}-\id_{[0,1)}$, in \cite{ZGJT01} in the continuous case
and in \cite{Moeser05} and \cite{Panayotaros05} in the discrete case:
given an initial condition $f$, the solutions of \eqref{eq:GT-time-version-1} and
\eqref{eq:NLS-gen-2} stay $\veps$ close -- measured in suitable Sobolev norms -- over long distances
$0\le t\le C/\veps$, see \cite{ZGJT01} and \cite{Moeser05,Panayotaros05} for the precise formulation.
Thus of special interest are stationary solutions of \eqref{eq:GT-time-version-1}, which can be
found making the ansatz
 \beq\label{eq:stationary}
    v_{\text{slow}}(t,x) = e^{i\omega t} f(x) ,
 \eeq
since they lead to breather like (quasi-)periodic solutions for the original equation
\eqref{eq:NLS-gen}, whose average  profile, for long $t\lesssim \veps^{-1}$, is given
by \eqref{eq:stationary}.
Before doing this it turns out to be advantageous to rewrite the non-local non-linear term in \eqref{eq:GT-time-version-1}:
we define a measure $\mu(B)$ by setting $\mu(B):= \tfrac{1}{2}\int_{-1}^1 \id_B(D(r))\, dr$
for any measurable set $B\subset \R$. Since $\mu(B)\ge 0$ and
$\mu(\R)= \int \id_\R(\tau)\mu(d\tau) = \int_0^1 \id_\R(D(r))\, dr = \int_0^1 dr=1$,
one sees that $\mu$ is a probability measure.
Moreover, as long as $d_0$ is locally integrable, $D$ is bounded and hence the probability
measure $\mu$ also has bounded support. Since $\mu$ is the image measure of normalized
Lebesgue measure on $[-1,1]$ under $D$, we can rewrite \eqref{eq:GT-time-version-1} as
    \beq\label{eq:GT-time-version-2}
        i\partial_t v_{\text{slow}}
        = d_{\text{av}}A v_{\text{slow}}
            - \int_\R
                U_{\tau}^{-1}\big[|U_{\tau} v_{\text{slow}}|^2 U_{\tau} v_{\text{slow}}\big]
              \, \mu(d\tau) .
    \eeq
The simplest case of dispersion management, $d_0=1$ on $[-1,0)$ and $d_0=-1$ on
$[0,1)$, i.e., $d_0= \id_{[-1,0)}-\id_{[0,1)}$, which is the case most
studied in the literature, corresponds to the measure $\mu$ having the density
$\id_{[0,1]}$, the uniform distribution on $[0,1]$.
More generally, if $d_0$ is piecewise
continuous on $[-1,1]$ and bounded away from zero, which is certainly a physically
reasonable assumption, or zero on an at most discrete subset of $[0,1]$, the inverse function
theorem, \cite{Strichartz00}, shows that $\mu$ has a density $\psi$, i.e,
$\mu(d\tau)= \psi(\tau)\, d\tau$, with $\psi$ given by
    \beq\label{eq:psi-def}
        \psi(\tau):= \sum_{t\in D^{-1}(\{\tau\})} |d_0(t)|^{-1}  ,
    \eeq
where $D^{-1}(\{\tau\})= \{t\in[0,1]\vert\, D(t)=\tau\}$.

Finally, we find it convenient to multi-linearize the non-linear and non-local term in \eqref{eq:GT-time-version-2} by introducing
 \beq\label{eq:Q}
    Q_\mu(v_1,v_2,v_3)(t)
    := \int_\R U_{\tau}^{-1}\big[ U_{\tau} v_1(t,\cdot) \ol{U_{\tau} v_2(t,\cdot)}U_{\tau} v_3(t,\cdot)\big]\, \mu(d\tau).
 \eeq
With this we rewrite \eqref{eq:GT-time-version-1} as
    \beq\label{eq:GT-time-version-3}
        i\partial_t v_{\text{slow}}
        = d_{\text{av}}A v_{\text{slow}}
            - Q_\mu(v_{\text{slow}},v_{\text{slow}},v_{\text{slow}}) .
    \eeq
The term $Q_\mu$ is closely related to the non-linear and non-local functionals appearing in the variational problems \eqref{eq:max-cont} and \eqref{eq:max-discr}. The ansatz \eqref{eq:stationary} in \eqref{eq:GT-time-version-3} yields the time independent Gabitov-Turitsyn equation; in our
notation,
 \beq\label{eq:GT}
    - \omega f
    = d_{\text{av}} A f - Q_\mu(f,f,f)  ,
 \eeq
which is a non-local non-linear eigenvalue equation for $f$. By testing \eqref{eq:GT} with suitable test
functions $g$ one arrives at the weak formulation
 \bdm
    -\omega \langle g, f\rangle
    =
    d_\text{av}\langle g, Af \rangle - \la g, Q_\mu(f,f,f)\ra
 \edm
where $\la g,f \ra$ is either the scalar product on $L^2(\R)$ given by
$\int \ol{g(x)} f(x)\, dx$ in the continuous case or the scalar product
$\sum_{x\in\Z} \ol{g(x)}f(x)$ on $l^2(\Z)$ in the discrete case. In the
continuous case we interpret the quadratic form $\langle g, Af \rangle$ as
$ \langle g, Af \rangle= \la \partial_x g, \partial_x f \ra = \langle g', f' \rangle $
by an integration by parts.
A formal calculation, using the unicity of $U_\tau$, yields
    \beq
        \la g, Q_\mu(f,f,f)\ra = \calQ_\mu^{\mathrm{c}/\mathrm{d}} (g,f,f,f)
    \eeq
where the four linear functional $\calQ_\mu^{\mathrm{c}/\mathrm{d}}$ is given by \eqref{eq:calQ-cont} in the continuous case and \eqref{eq:calQ-discr} in the discrete case, respectively.
The formal calculation yielding $\calQ_\mu^{\mathrm{c}/\mathrm{d}}$ is justified in the continuous and
discrete case using Lemma \ref{lem:boundedness-cont}, respectively Lemma \ref{lem:boundedness-discr}.
Thus the weak formulation of \eqref{eq:GT} is
    \beq\label{eq:GT-weak}
        -\omega \langle g, f\rangle
        =
        d_\text{av}\langle g, Af \rangle - \calQ_\mu^{\mathrm{c}/\mathrm{d}}(g,f,f,f) ,
    \eeq
supposed to hold for all $g\in l^2(\Z)$ in the discrete case and for any $g$ in the Sobolev
space $H^1(\R)$ in the continuous case.

Equation \eqref{eq:GT-weak} is the weak form of the Euler-\-Lagrange equation associated
with the energy
    \beq\label{eq:energy-gen}
        H(f) := \frac{d_{\text{av}}}{2} \la f, Af\ra - \frac{1}{4}\calQ_\mu^{\mathrm{c}/\mathrm{d}}(f,f,f,f) .
    \eeq
In particular, any minimizer of the associated constraint minimization problem
    \beq\label{eq:min-gen}
        M_\lambda^{d_\text{av}}:= \inf(H(f)\vert\, f\in X, \|f\|_2^2=\lambda) ,
    \eeq
where $X=l^2(\Z)$ in the discrete case and the Sobolev space $H^1(\R)$ in the continuous case,
will be, up to some minor technicalities, a solution of \eqref{eq:GT-weak} for some choice of Lagrange multiplier $\omega$, as long as the variational problem \eqref{eq:min-gen} admits minimizers.

The case of vanishing average dispersion/diffraction, $d_{\text{av}}=0$, is of particular
practical importance for applications, \cite{ESMBA98,ESMA00,Tetal03}, since in this case the
positive and negative dispersion/diffraction exactly cancel out. In the limit $d_\text{av}\to 0$,
the variational problem \eqref{eq:min-gen} yields, up to a minor sign change, the two
restricted variational problems \eqref{eq:max-cont}, respectively \eqref{eq:max-discr}, in the continuous, respectively discrete, case. Associated with these limiting two variational problems are the
Euler-Lagrange equations
    \beq\label{eq:GT-weak-zero-cont}
        \omega \la g,f\ra = \calQ_\mu^{\mathrm{c}}(g,f,f,f),  \quad \text{ for all } g\in L^2(\R)
    \eeq
    and
    \beq\label{eq:GT-weak-zero-discr}
        \omega \la g,f\ra = \calQ_\mu^{\mathrm{d}}(g,f,f,f),  \quad \text{ for all } g\in l^2(\Z)
    \eeq
which, in the language of differential equations, are the \emph{singular limits} of
\eqref{eq:GT-weak} for $d_{\text{av}}=0$  in the continuous, respectively discrete, case.
Note that the continuous version of \eqref{eq:GT-time-version-3} is elliptic for $d_\text{av}>0$,
whereas its singular limit \eqref{eq:GT-weak-zero-cont} is no longer elliptic.
This corresponds to the domain of the variational problem \eqref{eq:max-cont} increasing from
the Sobolev space $H^1(\R)$ to the full space $L^2(\R)$, due to the loss of second order derivatives.

Solutions of \eqref{eq:GT-weak-zero-cont} are precisely the dispersion management solitons for vanishing average dispersion and solutions of \eqref{eq:GT-weak-zero-discr} the diffraction management solitons for vanishing average diffraction. Theorems \ref{thm:existence-cont} and
\ref{thm:existence-discr} show that these solitons exist under rather mild conditions on $\mu$,
translating to very general and non-restrictive conditions on the profile $d_0$ for existence of dispersion management solitons, see Remark \ref{rem:conditions}.

As a final remark, we would like to note that even for $d_\text{av}>0$ existence of minimizers of
\eqref{eq:min-gen} has only been established for the special choice of profile
$d_0= \id_{[-1,0)}- \id_{[0,1)}$, corresponding to $\mu$ having density $\id_{[0,1]}$ in
the continuous case. The continuous case was done in \cite{ZGJT01} again using Lions'
concentration compactness principle.
Due to the absence of scaling in $l^2(\Z)$ there is a threshold phenomena for the existence of
minimizers of \eqref{eq:min-gen} in the discrete case similar to \cite{Weinstein99} which makes
the problem slightly harder than the continuous case:
The infimum  $M_\lambda^\text{diff}$ is  negative  only for sufficiently large
$\lambda$ and minimizers for \eqref{eq:min-gen} exist in the discrete case only if $\lambda$ is
large enough, depending on $d_\text{av}>0$, see \cite{Moeser05} and \cite{Panayotaros05} where
the existence of minimizers for piecewise continuous diffraction profile $d_0$ was shown using
a discrete version of Lions' concentration compactness principle.

In all above cases, using by now well-known arguments, see \cite{Cazenave03,CaLi82, Weinstein-86},
this variational approach to the existence of solutions of \eqref{eq:GT-weak} also shows that minimizers
of \eqref{eq:min-gen} for $d_\text{av}>0$, respectively maximizers of \eqref{eq:max-cont} and \eqref{eq:max-discr} for $d_\text{av}=0$ lead to orbitally stable solutions of \eqref{eq:GT-time-version-3}.

\section{The existence proof} \label{sec:existence}
We want to show that one can suitably massage any maximizing sequence for the variational problem \eqref{eq:max-cont}, respectively \eqref{eq:max-discr}, with translations and boosts of $L^2(\R)$, respectively translations of $l^2(\Z)$,
so that it lies in a compact subspace of $L^2$, respectively $l^2$.
The following lemma is the key result for this. First we need one more piece of notation. For $x> 0$ and $\alpha\in\R$ define
 \beq\label{eq:G}
    G_\alpha(x):= \Big(\big(x+\alpha^2\big)^{1/2} -\alpha\Big)^{-1/2} .
 \eeq
Note that $G_\alpha$ is a decreasing function on $\R_+$ which vanishes at infinity.
Moreover, for $z\in \R$ let $z_+:= \max(z,0)$.

\begin{lem}[Continuous case]\label{lem:fat-tail-bound-cont}
Let $\mu$ have a density $\psi$ satisfying
$0\le \psi\in L^2(\R)\cap L^4(\R)\cap L^4(\R, t^2dt)$.
Then there exists a constant $C$ depending only on
$\|\psi\|_{L^2(\R)}, \|\psi\|_{L^4(\R)}$ and
$\|\psi\|_{L^4(\R, t^2dt)}$ such that if
$f\in L^2(\R)$, $0< \veps <\|f\|_{L^2}$,
and $a,b\in\R$, with
 \beq\label{eq:fat-tail-cont}
   \int_{-\infty}^a |f(x)|^2\, dx \ge \frac{\veps^2}{2} \text{ and }
   \int_b^\infty |f(x)|^2\, dx    \ge \frac{\veps^2}{2} ,
 \eeq
 then
 \beq\label{eq:fat-tail-bound-cont}
   \calQ_\mu^\mathrm{c}(f,f,f,f) \le   P_1^c(\|f\|_{L^2}^4 -\veps^4/2) + C \|f\|_{L^2}^4 G_{1/2}((b-a)_+) .
 \eeq
 Moreover, whenever $c,d\in\R$, are such that
 \beq\label{eq:fourier-fat-tail-cont}
   \int_{-\infty}^c |\hatt{f}(k)|^2\, dk \ge \frac{\veps^2}{2} \text{ and }
   \int_d^\infty |\hatt{f}(k)|^2\, dk    \ge \frac{\veps^2}{2},
 \eeq
 then also
 \beq\label{eq:fourier-fat-tail-bound-cont}
   \calQ_\mu^\mathrm{c}(\hatt{f},\hatt{f},\hatt{f},\hatt{f})
   \le  P_1^\mathrm{c}(\|f\|_{L^2}^4 -\veps^4/2) +C \|f\|_{L^2}^4 G_{1/2}((d-c)_+) .
 \eeq
\end{lem}
We have a similar bound in the discrete case under much weaker assumptions on the measure $\mu$.

\begin{lem}[Discrete case]\label{lem:fat-tail-bound-discr}
Assume that $\mu$ is a bounded measure with bounded support. Then there
is a constant $C$ such that if $f\in l^2(\Z)$, $0<\veps <\|f\|_2$ and
$a,b\in\Z$, with
 \beq
 \begin{split} \label{eq:fat-tail-discr}
   \sum_{x < a} |f(x)|^2 \ge \frac{\veps^2}{2} \text{ and }
   \sum_{x > b} |f(x)|^2 \ge \frac{\veps^2}{2} ,
 \end{split}
 \eeq
 then
 \beq\label{eq:fat-tail-bound-discr}
   \calQ_\mu^\mathrm{d}\,(f,f,f,f) \le   P_1^\mathrm{d}(\|f\|_{l^2}^4 -\veps^4/2)
                    +C \|f\|_{l^2}^4 G_{1}((b-a+1)_+)
 \eeq
\end{lem}

\begin{remark}
Of course, by the monotonicity of $G_\alpha$, the bounds \eqref{eq:fat-tail-bound-cont}, respectively \eqref{eq:fourier-fat-tail-cont}, are strongest if one chooses the smallest $a$, respectively $c$,
and the largest $b$, respectively $d$,
under the restrictions \eqref{eq:fat-tail-cont}, respectively \eqref{eq:fourier-fat-tail-cont}.
Similarly, the bound \eqref{eq:fat-tail-bound-discr} is strongest if  one chooses
the smallest $a$ and the largest $b$ obeying \eqref{eq:fat-tail-discr}.
The choice $a\ge b$ or $c\ge d$ is allowed, although in this case the bounds do not yield
any information.
Unlike the continuum case, where for $f\in L^2(\R)$ and $0< \veps <\|f\|_{L^2}$ one can
always find $a<b$ and $c<d$ such that the bounds \eqref{eq:fat-tail-cont} and
\eqref{eq:fourier-fat-tail-cont} hold, it can happen in the discrete case that for some $f\in l^2(\Z)$
and  $0<\veps<\|f\|_{l^2}$ one has $a\ge b$ or even $a>b$ for any $a,b$ obeying
\eqref{eq:fat-tail-discr}. For example, this happens, even for all
$0<\veps<\|f\|_{l^2}$, if $f\in l^2(\Z)$ is concentrated on a single point in $\Z$. \\[0.1em]
Thus the bounds provided by Lemma \ref{lem:fat-tail-bound-cont} and
\ref{lem:fat-tail-bound-discr} yield the least information for strongly localized functions $f$, respectively their Fourier transforms $\hat{f}$.
At first this might seem counterintuitive since we intend to use these bounds in order to get concentration bounds on $f$, respectively $\hat{f}$, see Propositions \ref{prop:max-seq-tight-cont} and \ref{prop:max-seq-tight-discr} below.
\end{remark}

\bpf[Proof of Lemma \ref{lem:fat-tail-bound-cont}]
We prove only \eqref{eq:fat-tail-bound-cont} since, given Lemma \ref{lem:4linear-refined-cont},
the proof of \eqref{eq:fourier-fat-tail-bound-cont} is nearly identical.
We will write $\|f\|$ for $\|f\|_{L^2}$ in the following.
Of course, since the right hand side of \eqref{eq:fat-tail-bound-cont} is infinite if
$a\ge b$, we can assume that $a<b$. Let  $a'$ and $b'$ arbitrary numbers with
$a\le a'<b'\le b$ and split $f$ into
    \beq\label{eq:f-split}
        f= f_{-1} + f_0 + f_1
    \eeq
where we set $f_{-1}= f\id_{(-\infty, a')}$, $f_0= f\id_{[a', b']}$,
and $f_1= f\id_{(b',\infty)}$.
We will choose suitable $a'$ and $b'$ soon. Obviously, $\|f_j\|\le \|f\|$ for $j=-1,0,1$.
Moreover, since \eqref{eq:fat-tail-cont} and $a'\ge a$ and $b'\le b$
we also have
    \beq\label{eq:eps-lower-bound}
        \|f_{-1}\|^2 \text{ and } \|f_1\|^2 \ge \frac{\veps^2}{2}.
    \eeq

In order to bound $\calQ_\mu^\mathrm{c}(f,f,f,f)$ we use its multi-linearity,
    \beq\label{eq:calQ-split-1}
    \begin{split}
        \calQ_\mu^\mathrm{c}(f,f,f,f)
        &=
        \calQ_\mu^\mathrm{c}(f_{-1}+f_0+f_1,f,f,f) \\
        &= \calQ_\mu^\mathrm{c}(f_{-1},f,f,f) +\calQ_\mu^\mathrm{c}(f_0,f,f,f) +\calQ_\mu^\mathrm{c}(f_1,f,f,f).
    \end{split}
    \eeq
The term containing $f_0$ is simply bounded by
    \beq
        |\calQ_\mu^\mathrm{c}(f_0,f,f,f)|\le P_1^\mathrm{c} \|f_0\| \|f\|^3
    \eeq
by Lemma \ref{lem:boundedness-cont}. The other terms we further split into
    \beq\label{eq:calQ-split-2}
    \begin{split}
        \calQ_\mu^\mathrm{c}(f_1,f,f,f)
        &=
        \calQ_\mu^\mathrm{c}(f_1,f_{-1}+f_0+f_1,f,f) \\
        &= \calQ_\mu^\mathrm{c}(f_1,f_{-1},f,f) +\calQ_\mu^\mathrm{c}(f_1,f_0,f,f) +\calQ_\mu^\mathrm{c}(f_1,f_1,f,f)
    \end{split}
    \eeq
with a similar expression for $\calQ_\mu^\mathrm{c}(f_{-1},f,f,f)$. Again
    \beq
        |\calQ_\mu^\mathrm{c}(f_1,f_0,f,f)| \le P_1^\mathrm{c}\|f_{1}\| \|f_0\| \|f\|^2 \le P_1^\mathrm{c} \|f_0\| \|f\|^3
    \eeq
for the term containing $f_0$. Since the supports of $f_{-1}$ and $f_1$ have at least
distance $b'-a'$, the refined multi-linear bound of Lemma \ref{lem:4linear-refined-cont}
gives estimate
    \beq\label{eq:calQ-split-2-offdiag}
        |\calQ_\mu^\mathrm{c}(f_1,f_{-1},f,f)|
        \lesssim \frac{\|f_1\|\|f_{-1}\|\|f\|^2}{(b'-a')^{1/2}}
            \le \frac{\|f\|^4}{(b'-a')^{1/2}}
    \eeq
for the first term in \eqref{eq:calQ-split-2}. Terms of the form
$\calQ_\mu^\mathrm{c}(f_{-1},f_0,f,f)$ and $\calQ_\mu^\mathrm{c}(f_{-1},f_1,f,f)$ are estimated the same way.

Continuing similarly for the terms
$\calQ_\mu^\mathrm{c}(f_1,f_1,f,f)$, respectively $\calQ_\mu^\mathrm{c}(f_{-1},f_{-1},f,f)$,  we see that
there exist a constant $C<\infty$ such that
    \begin{align}
        \calQ_\mu^\mathrm{c}(f,f,f,f)
        &\le \calQ_\mu^\mathrm{c}(f_{-1},f_{-1},f_{-1},f_{-1}) + \calQ_\mu^\mathrm{c}(f_1,f_1,f_1,f_1) \nonumber \\
            &\phantom{\le~} + C \Big[ \|f_0\|\|f\|^3  + \frac{\|f\|^4}{(b'-a')^{1/2}}\Big]
                \nonumber \\
        &\le P_1^\mathrm{c}(\|f_{-1}\|^4 + \|f_1\|^4)
                + C \Big[ \|f_0\|\|f\|^3  + \frac{\|f\|^4}{(b'-a')^{1/2}}\Big] , \label{eq:near-max-1}
    \end{align}
where the second inequality follows again from Lemma \ref{lem:boundedness-cont}.
Using \eqref{eq:eps-lower-bound}, we get
    \bdm
        \|f_{-1}\|^4 + \|f_1\|^4
            = (\|f_{-1}\|^2 + \|f_1\|^2)^2 - 2\|f_{-1}\|^2\|f_1\|^2
            \le \|f\|^4 - \frac{\veps^4}{2}
    \edm
since $\|f_{-1}\|^2 + \|f_1\|^2 \le \|f\|^2$ always. In particular, \eqref{eq:near-max-1} gives
    \beq\label{eq:near-max-2}
        \calQ_\mu^\mathrm{c}(f,f,f,f)
        \le
        P_1^\mathrm{c}\big( \|f\|^4 - \frac{\veps^4}{2} \big) + C \Big[ \|f_0\|\|f\|^3  + \frac{\|f\|^4}{(b'-a')^{1/2}}\Big] .
    \eeq
To choose $a'$ and $b'$ let  $0<l<b-a$ and note that
    \beq\label{eq:pidgeonholing}
    \begin{split}
        \int_a^{b-l} \int_\eta^{\eta+l} |f(x)|^2\, dx\, d\eta
            &= \int_{\substack{a\le \eta\le b-l\\ \eta\le x \le \eta+l}} |f(x)|^2\, dx\, d\eta
            \le \int_{\substack{a\le x \le b\\ x-l \le \eta \le x}} |f(x)|^2\, dx\, d\eta \\
            &= l \int_a^b |f(x)|^2\, dx \le l \|f\|^2 .
    \end{split}
    \eeq
By the mean value theorem and \eqref{eq:pidgeonholing}, there exists
$\eta'\in(a,b-l)$ such that
    \bdm
        (b-a-l) \int_{\eta'}^{\eta'+l} |f(x)|^2\, dx
            = \int_a^{b-l} \int_\eta^{\eta+l} |f(x)|^2\, dx\, d\eta.
    \edm
Thus, with the choice $a'= \eta'$ and $b'= \eta'+l$ we have $b'-a'=l$,  $a<a'<b'<b$, and
    \bdm
        \|f_0\|^2 \le \frac{l}{b-a-l} \|f\|^2.
    \edm
Plugging this into \eqref{eq:near-max-2} yields
    \beq\label{eq:near-max-3}
        \calQ_\mu^\mathrm{c}(f,f,f,f)
        \le
        P_1^\mathrm{c}( \|f\|^4 - \frac{\veps^4}{2} )
        + C \|f\|^4 \Big[ \big( \frac{l}{b-a-l} \big)^{1/2}  + \frac{1}{l^{1/2}}\Big]
    \eeq
for any $0<\veps<\|f\|$ and all $0<l<b-a$. The choice $l= \sqrt{b-a+1/4}-1/2$, which is
allowed since $0<\sqrt{s+1/4}-1/2 < s$ for any $s> 0$, gives
    \bdm
        \frac{l}{b-a-l}  = \frac{1}{l}.
    \edm
Hence \eqref{eq:near-max-3} yields \eqref{eq:fat-tail-bound-cont}.
\epf
\bpf[Proof of Lemma \ref{lem:fat-tail-bound-discr}]
As in the continuous case, we can assume that $a,b\in\Z$ with $a\le b$ and we write $\|f\|$ for
the $l^2$-norm of a function $f\in l^2(\Z)$.
Again, for any choice $a\le a^\prime\le b^\prime\le b$ one puts
$f_{-1}= f\id_{(-\infty,a^\prime)}=f\id_{(-\infty, a^\prime-1]}$,
$f_1=f\id_{(b^\prime,\infty)}= f\id_{[b^\prime+1,\infty)}$, and
$f_0=f\id_{[a^\prime,b^\prime]}$. Furthermore, let $l$ be the number of points in
$[a^\prime,b^\prime]$, i.e., $l=b^\prime-a^\prime + 1$ and note that
$\dist(\supp(f_{-1}),\supp(f_1))= l+1$. Then we argue exactly as in
the continuous case, but use the bounds from Lemma \ref{lem:boundedness-discr} and
\ref{lem:4linear-refined-discr} instead, to see that
    \beq\label{eq:near-max-2-discr}
        \calQ_\mu^\mathrm{d}(f,f,f,f)
        \le
        P^d_1( \|f\|_2^4 - \frac{\veps^4}{2} )
            + C \Big[ \|f_0\|_2\|f\|_2^3  + \|f\|_2^4 (l+1)^{-(l+1)/4}\Big]
    \eeq
holds. The discrete version of \ref{eq:pidgeonholing} now reads
    \beq \label{eq:pidgeonholing-discr}
        \begin{split}
        \sum_{\eta= a}^{b-l+1} \sum_{x=\eta}^{\eta+l-1} |f(x)|^2
            \le \sum_{x= a}^{b} \sum_{\eta= x-l+1}^{x} |f(x)|^2
            \le  l \|f\|_2^2 .
    \end{split}
    \eeq
By pidgeonholing, since the number of points in $[a,b-l+1]$ is $b-a+2-l$, there must exists $\eta^\prime$ with
$a\le \eta^\prime \le b-l+1$ and
    \bdm
        (b-a+2-l) \sum_{x= \eta^\prime}^{\eta^\prime+l-1} |f(x)|^2
            \le \sum_{\eta= a}^{b-l+1} \sum_{x=\eta}^{\eta+l-1} |f(x)|^2  .
    \edm
Thus, choosing $a^\prime=\eta^\prime$ and $b^\prime=\eta^\prime+l-1$, the bounds
\eqref{eq:near-max-2-discr} and \eqref{eq:pidgeonholing-discr} give
    \beq\label{eq:near-max-3-discr}
        \calQ_\mu^\mathrm{d}(f,f,f,f)
        \le
        P^d_1( \|f\|_2^4 - \frac{\veps^4}{2} )
            + C \|f\|_2^4 \Big[ \big( \frac{l}{b-a+2-l} \big)^{1/2}  +  (l+1)^{-(l+1)/4} \Big]
    \eeq
Now there exists $l\in\N$ with $l\le (b-a+2)^{1/2} <l+1$, which is an allowed choice for $l$, i.e.,
it obeys $1\le l\le b-a+1$ for any $a\le b \in\Z $. With this choice the estimates
    \bdm
        \big( \frac{l}{b-a+2-l} \big)^{1/2}
            \le \big( \frac{(b-a+2)^{1/2}}{b-a+2-(b-a+2)^{1/2}} \big)^{1/2}
            =   \big( (b-a+2)^{1/2}-1 \big)^{-1/2}
    \edm
and, since $l\ge 1$,
    \bdm
        (l+1)^{-(l+1)/4} \le (l+1)^{-1/2} \le \big( b-a+2\big)^{-1/4}
        \le \big( (b-a+2)^{1/2}-1\big)^{-1/2}
    \edm
show that \eqref{eq:near-max-3-discr} implies \eqref{eq:fat-tail-bound-discr}.
\epf

Lemma \ref{lem:fat-tail-bound-cont} has strong consequences for maximizing sequences of the
variational problem \eqref{eq:max-cont}. Recall that $(f_n)_n\subset L^2(\R)$ is a maximizing
sequence for \eqref{eq:max-cont} if $\|f_n\|_{L^2}^2=\lambda>0$ for all $n\in\N$ and
    \bdm
        \lim_{n\to\infty} \calQ_\mu^\mathrm{c}(f_n,f_n,f_n,f_n)
        = 
            \sup \Big( \calQ_\mu^\mathrm{c}(g,g,g,g)\big\vert\, g\in L^2(\R), \|g\|_{L^2}^2=\lambda \Big) .
    \edm
The following proposition shows that any maximizing sequence for the variational problem
\eqref{eq:max-cont} is \emph{tight modulo translations and shifts}.

\begin{prop}[Tightness, continuous case] \label{prop:max-seq-tight-cont}
Let $(f_n)_{n\in\N}\subset L^2(\R)$ be a maximizing sequence for the variational problem
\eqref{eq:max-cont} with $\lambda=\|f_n\|_2^2 >0$. Then there exist shifts $\xi_n$ and boosts
$v_n$ such that
 \beq\label{eq:max-seq-tight}
  \lim_{R\to\infty}\sup_{n\in\N} \int_{|x -\xi_n|>R} |f_n(x)|^2\, dx = 0 ,
 \eeq
 and
 \beq\label{eq:max-seq-tight-fourier}
  \lim_{L\to\infty}\sup_{n\to\infty} \int_{|k -v_n|>L} |\hatt{f}_n(k)|^2\, dk = 0 .
 \eeq
\end{prop}
\bpf We will only  prove \eqref{eq:max-seq-tight} since, given the bound
\eqref{eq:fourier-fat-tail-bound-cont} in Lemma \ref{lem:fat-tail-bound-cont},
the proof of \eqref{eq:max-seq-tight-fourier} is identical.
To prove \eqref{eq:max-seq-tight},
we have to show that there exist  shifts $\xi_n$ such that for any $0<\veps<\sqrt{\lambda}$ there
exists $R_\veps<\infty$  with
 \beq\label{eq:max-seq-veps}
  \sup_{n\in\N} \int_{|x -\xi_n|>R_\veps} |f_n(x)|^2\, dx \le \veps^2 .
 \eeq
Define
 \beq\label{eq:a-n-eps-cont}
   a_{n,\veps} := \inf\big( a\in\R: \int_{-\infty}^{a} |f_n(x)|^2\, dx  \ge \frac{\veps^2}{2} \big) -1
 \eeq
 and
 \beq\label{eq:b-n-eps-cont}
   b_{n,\veps} :=  \sup\big( b\in\R: \int_{b}^\infty |f_n(x)|^2\, dx  \ge \frac{\veps^2}{2} \big) +1 .
 \eeq
Both $a_{n,\veps}$ and $ b_{n,\veps}$ exist and are finite since $f_n\in L^2(\R)$ and
$f_n\not\equiv 0$.
Moreover, $a_{n,\veps}<b_{n,\veps}$ for all $n\in\N$ and $0<\veps<\sqrt{\lambda}$,
and they are monotone: $a_{n,\veps_1}\le a_{n,\veps_2}$ and $b_{n,\veps_1}\ge b_{n,\veps_2}$
for all $0<\veps_1\le \veps_2<\sqrt{\lambda}$ and $n\in\N$.
Fix $0<\veps_0<\sqrt{\lambda}$ and put
    \beq\label{eq:choice-xi-n}
        \xi_n:= (b_{n,\veps_0} + a_{n,\veps_0})/2,
    \eeq
or choose any point in $[a_{n,\veps_0},b_{n,\veps_0}]$,  and assume, for now,
that
    \beq\label{eq:choice-R-eps}
        R_\veps:= \sup_{n\in\N} (b_{n,\veps}-a_{n,\veps}) <\infty
    \eeq
for $0<\veps\le \veps_0$ and put $R_\veps= R_{\veps_0}$ for $\veps_0<\veps<\sqrt{\lambda}$.
With this and \eqref{eq:a-n-eps-cont} and \eqref{eq:b-n-eps-cont} one can easily check that
    \bdm
        \int_{|x -\xi_n|>R_\veps} |f_n(x)|^2 \, dx
        \le \int_{-\infty}^{a_{n,\veps}} |f_n(x)|^2\, dx + \int_{b_{n,\veps}}^\infty |f_n(x)|^2\, dx
        \le \veps^2
    \edm
for all $n\in\N$, which proves \eqref{eq:max-seq-veps} if we can show that
$R_\veps$ defined in \eqref{eq:choice-R-eps} is indeed finite.
This is where the bound \eqref{eq:fat-tail-bound-cont} of Lemma \ref{lem:fat-tail-bound-cont} enters.

By our choice of $a_{n,\veps}$ and $b_{n,\veps}$ in \eqref{eq:a-n-eps-cont} and \eqref{eq:b-n-eps-cont}
we have
    \bdm
        \int_{-\infty}^{a_{n,\veps}+2} |f_n(x)|^2\, dx \ge \frac{\veps^2}{2} \quad\text{ and }\quad
        \int_{b_{n,\veps}-2}^{\infty} |f_n(x)|^2\, dx \ge \frac{\veps^2}{2}.
    \edm
Thus with $a= a_{n,\veps}+2$ and $b= b_{n,\veps}-2$ the assumption \eqref{eq:fat-tail-cont} of Lemma \ref{lem:fat-tail-bound-cont} is fulfilled. Putting $R_{n,\veps}= b_{n,\veps}- a_{n,\veps}$,
using the scaling $P_1^\mathrm{c}\|f_n\|_2^4 = P_1^\mathrm{c}\lambda^2= P_\lambda$, see \eqref{eq:scaling} in
the appendix, and rearranging \eqref{eq:fat-tail-bound-cont} a bit, yields
 \beq\label{eq:rearranged}
    P_1^\mathrm{c} \frac{\veps^4}{2} +\calQ_\mu^\mathrm{c}(f_n,f_n,f_n,f_n) - P_\lambda
    \le C \lambda^2 \, G_{1/2}( (R_{n,\veps}-4)_+ ) .
 \eeq
 Since $f_n$ is a maximizing sequence for $P_\lambda$ we have
 $\lim_{n\to\infty} \calQ_\mu^\mathrm{c}(f_n,f_n,f_n,f_n)= P_\lambda$. The function
 $G_{1/2}$ is decreasing on $\R_+$.
 Thus
    \bdm
        \liminf_{n\to\infty} G_{1/2}( (R_{n,\veps}-4)_+ ) = G_{1/2}((\limsup_{n\to\infty}R_{n,\veps}-4)_+).
    \edm
 Hence taking the limit $n\to\infty$ in \eqref{eq:rearranged} one sees
 \beq\label{eq:bound}
   0< P_1^\mathrm{c} \frac{\veps^4}{2}
   \le C \lambda^2 \liminf_{n\to\infty} G_{1/2}( (R_{n,\veps}-4)_+ )
    = C \lambda^2 G_{1/2}((\limsup_{n\to\infty}R_{n,\veps}-4)_+) .
 \eeq
 Since $G_{1/2}$ goes to zero at infinity, the bound \eqref{eq:bound} shows that
 \beq
   \limsup_{n\to\infty}R_{n,\veps} <\infty \quad \text{for all } 0<\veps<\sqrt{\lambda}
 \eeq
 which proves \eqref{eq:choice-R-eps} and hence the Lemma.
\epf
Of course, we have an analogous proposition for any maximizing sequence of the discrete
variational problem \eqref{eq:max-discr}.

\begin{prop}[Discrete case] \label{prop:max-seq-tight-discr}
Let $(f_n)_{n\in\N}\subset l^2(\Z)$ be a maximizing sequence for the variational problem
\eqref{eq:max-discr} with $\lambda=\|f_n\|_2^2$. Then there exist shifts $\xi_n\in\Z$
 \beq\label{eq:max-seq-tight-discr}
  \lim_{R\to\infty}\sup_{n\in\N} \sum_{|x -\xi_n|>R} |f_n(x)|^2 = 0
 \eeq
\end{prop}
\bpf
Given Lemma \ref{lem:fat-tail-bound-discr}, the proof of Proposition \ref{prop:max-seq-tight-discr}
is virtually identical to the proof of Proposition \ref{prop:max-seq-tight-cont}.
\epf
Propositions \ref{prop:max-seq-tight-cont} and \ref{prop:max-seq-tight-discr} are
key to our proof of the existence of maximizers.
\bpf[Proof of Theorem \ref{thm:existence-cont} and \ref{thm:existence-discr}:]
We only give the details in the continuous case. The discrete case follows by a similar reasoning.

The idea is to use Proposition \ref{prop:max-seq-tight-cont} and Lemma
\ref{lem:strong-convergence-L2} in order to massage an arbitrary
maximizing sequence into a strongly convergent sequence. Its
strong limit will then furnish the sought after maximizer.

Let $(f_n)_n\subset L^2(\R)$ be an arbitrary
maximizing sequence of the variational problem \eqref{eq:max-cont}.
Proposition \ref{prop:max-seq-tight-cont} guarantees the existence of shifts
$\xi_n\in\R$ and boosts $v_n\in\R$ such that \eqref{eq:max-seq-tight} and
\eqref{eq:max-seq-tight-fourier} hold.
Define the shifted and boosted sequence
    \bdm
        \tilde{f}_n(x):= \big(e^{ixv_n}e^{-i\xi_n P}f_n\big)(x) =e^{iv_n} f_n(x-\xi_n).
    \edm
where $P=-i\partial_x$ is the one-dimensional momentum operator.

Note that $\|\tilde{f}_n\|_2^2= \|f_n\|_2^2=\lambda$ since shifts and boost are unitary
operations on $L^2(\R)$.
As discussed in the introduction, due to the Galilei covariance of the free
Schr\"odinger equation, see \eqref{eq:Galilei-trafo},  the functional $\calQ_\mu^\mathrm{c}$ is
invariant under shifts and boosts, i.e.,
$\calQ_\mu^\mathrm{c}(\tilde{f}_n,\tilde{f}_n,\tilde{f}_n,\tilde{f}_n)= \calQ_\mu^\mathrm{c}(f,f,f,f)$.
Hence $(\tilde{f}_n)_n$ is also a maximizing sequence.

Certainly $|\wti{f}_n(x)|= |f_n(x-\xi_n)|$ for all $n\in\N$. The Fourier transform of $\wti{f}_n$ is given by
 \beq
    \hatt{\wti{f}_n}(k)= \frac{1}{\sqrt{2\pi}} \int e^{-ixk} e^{ixv_n} f_n(x-\xi_n)\, dx
    = e^{-i\xi_n(k-v_n)} \hatt{f}_n(k-v_n).
 \eeq
Thus also $|\hatt{\wti{f}_n}(k)|= |\hatt{f}_n(x-v_n)|$.  In particular,
\eqref{eq:max-seq-tight} and \eqref{eq:max-seq-tight-fourier} show that the maximizing
sequence $(\tilde{f}_n)_n$  is tight in the sense of Lemma \ref{lem:strong-convergence-L2}.

Since $(\tilde{f}_n)_n$ is bounded in $L^2(\R)$, the weak compactness of the unit ball,
\cite{LiebLoss}, guarantees the existence of a weakly converging subsequence
$(\tilde{f}_{n_j})_j$ of $(f_n)_n$. Obviously, this subsequence is also tight in the
sense of  Lemma \ref{lem:strong-convergence-L2} and hence converges even strongly in $L^2$.
We set
    \bdm
        f= \lim_{j\to\infty} \tilde{f}_{n_j} .
    \edm
By strong convergence $ \|f\|_2^2 = \lim_{j\to\infty} \|\tilde{f}_{n_j}\|_2^2 = \lambda $.
To conclude that $f$ is the sought after maximizer we note that by the following Lemma
\ref{lem:lipshitz-continuity} the map $f\mapsto \calQ_\mu^\mathrm{c}(f,f,f,f)$ is continuous on $L^2(\R)$.
Hence
    \bdm
        \calQ_\mu^\mathrm{c}(f,f,f,f) = \lim_{j\to\infty}\calQ_\mu^\mathrm{c}(\tilde{f}_{n_j},\tilde{f}_{n_j},\tilde{f}_{n_j},\tilde{f}_{n_j})
        = P_\lambda .
    \edm
where the last equality follows since $(\tilde{f}_{n})_n$ is a maximizing sequence.
Thus $f$ is a maximizer for the variational problem \eqref{eq:max-cont}.

The proof that the above maximizer is a weak solution of the associated Euler--Lagrange equation
\eqref{eq:GT-weak-zero-cont} is standard in the calculus of variations, we sketch it for the
convenience of the reader: Let $\varphi(f)=\varphi_{\text{cont}}(f)= \calQ_\mu^\mathrm{c}(f,f,f,f)$.
Lemma \ref{lem:derivative} below shows that the derivative of the functional
$\varphi$ at any $f\in L^2(\R)$ is given by the linear map $D\varphi(f)[h]= 4\re\calQ_\mu^\mathrm{c}(h,f,f,f)$.
Similarly, one can check that the derivative of $\psi(f)=\|f\|_2^2= \la f,f\ra$ is given by
$D\psi(f)[h]= 2\re\la h,f\ra$.
Note that although, in our convention for the inner product, the map $h\mapsto \la h,f\ra$ is
anti-linear, the map $h\mapsto \re\la h,f\ra$ is linear. Similarly, one easily checks that
for fixed $f$ the map $h\mapsto \re\calQ_\mu^\mathrm{c}(h,f,f,f)$ is linear.

Now let $f$ be any maximizer of the constraint variational problem \eqref{eq:max-cont} and
$h\in L^2(\R)$ arbitrary. Define, for any $(s,t)\in\R^2$,
 \begin{align*}
    F(s,t)&:= \varphi(f+ sf +t h) , \\
    G(s,t)&:= \psi(f+ sf +t h ) .
 \end{align*}
Note that
 \begin{align*}
    \nabla F(s,t)&= \left(\begin{array}{c} D\varphi(f+ sf +t h)[f] \\ D\varphi(f+ sf +t h)[h] \end{array} \right)\\
    &= 4\left(\begin{array}{c} \re\calQ_\mu^\mathrm{c}(f,f+ sf +t h,f+ sf +t h,f+ sf +t h) \\
                              \re\calQ_\mu^\mathrm{c}(h,f+ sf +t h,f+ sf +t h,f+ sf +t h)  \end{array} \right)
 \end{align*}
and
 \bdm
    \nabla G(s,t)= \left(\begin{array}{c} D\psi(f+ sf +t h)[f] \\ D\psi(f+ sf +t h)[h] \end{array} \right)
    = 2\left(\begin{array}{c} \re\la f,f+ sf +t h\ra \\
                              \re \la h,f+ sf +t h\ra  \end{array} \right) .
 \edm
Since $\la f,f\ra = \lambda\not= 0$,
 \bdm
    \nabla G(0,0)= 2\left(\begin{array}{c} \la f,f\ra \\
                              \re \la h,f\ra  \end{array} \right)
 \edm
 is not the zero vector in $\R^2$ and since $\nabla G(s,t)$ depends multi-linearly, in particular
 continuously, on $(s,t)$, the implicit function theorem \cite{Strichartz00} shows that there
 exists an open interval $I\subset\R$ containing $0$ and a differentiable function $\phi$ on $I$
 with $\phi(0)=0$ such that
 \bdm
    \lambda= \|f\|_2^2 = G(0,0) = G(\phi(t),t)
 \edm
 for all $t\in I$. Consider the function $I\ni t\mapsto F(\phi(t),t)$. Since $f$ is a maximizer
 for the constraint variational problem \eqref{eq:max-cont}, $F(\phi(t),t)$ has a local maximum at
 $t=0$. Hence, using the chain rule,
 \bdm
    0 = \frac{d F(\phi(t),t)}{dt}\Big\vert_{t=0}
        = \nabla F(0,0)\cdot\left(\begin{array}{c}\phi'(0)\\1\end{array}\right)
        = 4 \calQ_\mu^\mathrm{c}(f,f,f,f) \phi'(0) + 4\re\calQ_\mu^\mathrm{c}(h,f,f,f).
 \edm
 Since $\lambda= G(\phi(t),t)$, the chain rule also yields
 \bdm
    0= \frac{d G(\phi(t),t)}{dt}\Big\vert_{t=0}
        = \nabla G(0,0)\cdot\left(\begin{array}{c}\phi'(0)\\1\end{array}\right)
        = 2 \la f,f\ra \phi'(0) + 2\re \la h,f\ra.
 \edm
 Solving this for $\phi'(0)$ and plugging it back into the expression for the derivative of $F$,
 we see that
 \bdm
    \frac{\calQ_\mu^\mathrm{c}(f,f,f,f)}{\la f,f\ra} \re \la h,f\ra = \re\calQ_\mu^\mathrm{c}(h,f,f,f).
 \edm
 In other words, with $\omega := \calQ_\mu^\mathrm{c}(f,f,f,f)/\la f,f\ra= P_\lambda/\lambda >0$ and $f$ any maximizer of
 \eqref{eq:max-cont}, we have
 \beq\label{eq:euler-lagrange-real}
     \re(\omega\la h,f\ra) = \re\calQ_\mu^\mathrm{c}(h,f,f,f)
 \eeq
 for any $h\in L^2(\R)$. Replacing $h$ by $ih$ in \eqref{eq:euler-lagrange-real}, one gets
 \beq\label{eq:euler-lagrange-imaginary}
     \im(\omega\la h,f\ra) = \im\calQ_\mu^\mathrm{c}(h,f,f,f)
 \eeq
 for all $h\in L^2(\R)$. \eqref{eq:euler-lagrange-real} and \eqref{eq:euler-lagrange-imaginary} together
 show
 \bdm
    \omega \la h,f\ra = \calQ_\mu^\mathrm{c}(h,f,f,f)
 \edm
 for any $h\in L^2(\R)$, that is, $f$ is a weak solution of the
 dispersion management equation \eqref{eq:GT-weak-zero-cont}.
\epf

\begin{lem}\label{lem:lipshitz-continuity}
\itemthm \label{lem:lipshitz-continuity-cont} If the measure $\mu$ has density $\psi\in L^2(\R)$
then  the map $L^2(\R)\ni f\mapsto \calQ_\mu^\mathrm{c}(f,f,f,f)$ is locally Lipshitz continuous on $L^2(\R)$.\\[0.2em]
\itemthm \label{lem:lipshitz-continuity-discr} If the measure $\mu$ is bounded then the
map $l^2(\Z)\ni f\mapsto \calQ_\mu^\mathrm{d}(f,f,f,f)$ is locally Lipshitz continuous on $l^2(\Z)$.
\end{lem}
\bpf
 Using the multi-linearity of $\calQ_\mu^\mathrm{c}$, given $f,g\in L^2(\R)$, one has
 \beq
 \begin{split}
    \varphi & (f)-\varphi(g) = \calQ_\mu^\mathrm{c}(f,f,f,f)- \calQ_\mu^\mathrm{c}(g,g,g,g) \\
    &= \calQ_\mu^\mathrm{c}(f-g,f,f,f) + \calQ_\mu^\mathrm{c}(g,f-g,f,f) + \calQ_\mu^\mathrm{c}(g,g,f-g,f) +\calQ_\mu^\mathrm{c}(g,g,g,f-g)
 \end{split}
 \eeq
 This together with the triangle inequality and the a-priori bound of Lemma \ref{lem:boundedness-cont}
 immediately yields
\beq
 \begin{split}
   |\varphi(f)-\varphi(f)|
    &\le
        P_1^\mathrm{c} \sum_{j=0}^3
                \|f\|_2^{3-j} \|f-g\|_2 \|g\|_2^{j} \\
    &\le 4 P_1^\mathrm{c} \max(1,\|f\|_2^3, \|g\|_2^3) \|f-g\|_2 .
 \end{split}
 \eeq
 The discrete case is proven the same way using Lemma \ref{lem:boundedness-discr}
\epf
In the proof of Theorems \ref{thm:existence-cont} and \ref{thm:existence-discr} we needed one
more  technical result, about the differentiability of the non-linear functionals
$\calQ_\mu^{\mathrm{c}/\mathrm{d}}$:
\begin{lem}\label{lem:derivative}
\itemthm \label{lem:derivative-cont} If the measure $\mu$ has density $\psi\in L^2(\R)$ then
 the map $L^2(\R)\ni f\mapsto\varphi_\mu^\mathrm{c}(f)=\calQ_\mu^\mathrm{c}(f,f,f,f)$ is continuously differentiable with derivative $D\varphi_\mu^\mathrm{c}(f)[h]= 4\re\calQ_\mu^\mathrm{c}(h,f,f,f)$.  \\[0.2em]
\itemthm \label{lem:derivative-discr} If the measure $\mu$ is bounded then the map
    $l^2(\Z)\ni f\mapsto\varphi_\mu^\mathrm{d}(f)=\calQ_\mu^\mathrm{d}(f,f,f,f)$ is continuously differentiable with derivative $D\varphi_\mu^\mathrm{d}(f)[h]= 4\re\calQ_\mu^\mathrm{d}(h,f,f,f)$.
\end{lem}
\bpf
Using the multi-linearity of $\calQ_\mu^\mathrm{c}$, one can check that for any $f,h\in L^2(\R)$
 \begin{multline}
    \varphi_\mu^\mathrm{c}(f+h) \\
    = \varphi_\mu^\mathrm{c}(f)
        + \calQ_\mu^\mathrm{c}(f,f,f,h)+\calQ_\mu^\mathrm{c}(f,f,h,f) + \calQ_\mu^\mathrm{c}(f,h,f,f) + \calQ_\mu^\mathrm{c}(h,f,f,f)
        + O(\|h\|_2^2) \\
    = \varphi_\mu^\mathrm{c}(f) + 4\re\calQ_\mu^\mathrm{c}(h,f,f,f) + O(\|h\|_2^2)
 \end{multline}
 where in the term $O(\|h\|_2^2)$ we gathered expressions of the form $\calQ_\mu^\mathrm{c}(h,f,f,h)$, and
 $\calQ_\mu^\mathrm{c}(h,h,f,f)$ or $\calQ_\mu^\mathrm{c}(h,h,f,f+h)$ and similar which, by the a-priori bound from Lemma \ref{lem:boundedness-cont},
 are bounded by $C\|h\|_2^2$ with $C\le P_1^\mathrm{c} \max(1,\|f\|_2^2,\|h\|_2^2)$. This shows that
 $\varphi_\mu^\mathrm{c}$ is differentiable with derivative
 $D\varphi_\mu^\mathrm{c}(f)[h]= 4\re\calQ_\mu^\mathrm{c}(h,f,f,f)$.
 Moreover,
 \bdm
 \begin{split}
    D\varphi_\mu^\mathrm{c}(f)[h] - & D\varphi_\mu^\mathrm{c}(g)[h]
     = 4\re\big(\calQ_\mu^\mathrm{c}(h,f,f,f) - \calQ_\mu^\mathrm{c}(h,g,g,g)\big) \\
    & = 4\re\big(\calQ_\mu^\mathrm{c}(h,f-g,f,f) + \calQ_\mu^\mathrm{c}(h,g,f-g,f) + \calQ_\mu^\mathrm{c}(h,g,g,f-g) \big)  .
 \end{split}
 \edm
 Hence using the bound from Lemma \ref{lem:boundedness-cont} again, we see
 \beq
  \sup_{\|h\|_2\le 1} \big| D\varphi_\mu^\mathrm{c}(f)[h] - D\varphi_\mu^\mathrm{c}(g)[h] \big|
   \lesssim (\|f\|_2^2 + \|f\|_2\|g\|_2 + \|g\|_2^2) \|f-g\|_2
 \eeq
 which shows that the derivative $D\varphi_\mu^\mathrm{c}$ is even locally Lipshitz continuous.
 The discrete case is proven analogously.
\epf

\vspace{5mm}

\appendix
\setcounter{section}{0}
\renewcommand{\thesection}{\Alph{section}}
\renewcommand{\theequation}{\thesection.\arabic{equation}}
\renewcommand{\thethm}{\thesection.\arabic{thm}}

\section{Strong convergence} \label{sec:strong-convergence}
A key step in our existence proof of maximizers of the variational problems \eqref{eq:max-cont}
and  \eqref{eq:max-discr} is the characterization of strong convergence in $L^2(\R)$ and
$l^2(\Z)$. We need only the respective one-dimensional versions, but give the result and
its proof for all dimensions. In the discrete case the simple characterization
of strong convergence extends to $l^p(\Z^d)$ for any $1\le p <\infty$. We start with

\begin{lem}\label{lem:strong-convergence-L2}
A sequence $(f_n)_{n\in\N}\subset L^2(\R^d)$ is strongly converging to $f$ in $L^2(\R^d)$
if and only if it is weakly convergent to $f$ and
 \begin{align}
    \lim_{R\to\infty} \limsup_{n\to\infty} \int_{|x|>R} |f_n(x)|^2 \, dx &= 0,
    \label{eq:tight-real}\\
    \lim_{L\to\infty} \limsup_{n\to\infty} \int_{|k|>L} |\hatt{f}_n(k)|^2 \, dk &= 0,
    \label{eq:tight-fourier}
 \end{align}
 where $\hatt{f}$ is the Fourier transform of $f$.
\end{lem}
\begin{remarks}
\itemthm A bounded sequence in $L^2(\R^d)$ can converge weakly to zero by vanishing, splitting,
    or oscillating to death. \eqref{eq:tight-real} prevents splitting and vanishing and
    \eqref{eq:tight-fourier} prevents oscillating to death, which corresponds to vanishing or
    splitting in Fourier space. \\[0.2em]
\itemthm \label{rem:equivalence} Given a sequence $(f_n)_{n\in\N}\subset L^2(\R^d)$ the bounds
\eqref{eq:tight-real} and \eqref{eq:tight-fourier} certainly hold if there exists functions
$H,F\ge 1$ with $\lim_{|x|\to\infty}H(x)=\infty= \lim_{|k|\to\infty}F(k)$ and
 \begin{align}
    \limsup_{n\to\infty} \int H(x) |f_n(x)|^2 \, dx &<\infty,
    \label{eq:tight-H}\\
    \limsup_{n\to\infty} \int F(k) |\hatt{f}_n(k)|^2 \, dk &<\infty,
    \label{eq:tight-F}
 \end{align}
 on the other hand, it is easy to see that once \eqref{eq:tight-real} holds then there exist a function
 $H$ bounded below by one and growing to infinity at infinity such that \eqref{eq:tight-H} holds. So
 \eqref{eq:tight-real} and \eqref{eq:tight-H}, and hence also \eqref{eq:tight-fourier} and \eqref{eq:tight-F} are equivalent. In particular, in the proof of the difficult part of Lemma \ref{lem:strong-convergence-L2} one could use Rellich's compactness result, see \cite{ReSi4}.
 We prefer, however, the proof given below, which uses only simple properties of compact
 operators. \\[0.2em]
\itemthm Of course Lemma \ref{lem:strong-convergence-L2} holds also with $\limsup_{n\to\infty}$
replaced by $\sup_{n\in\N}$.
\end{remarks}
\bpf Assume that $f_n$ converges to $f$ strongly. Then it certainly converges weakly to $f$, that is, $\lim_{n\to\infty}\la\varphi, f_n\ra =  \la \varphi, f\ra$ for all $\varphi\in L^2(\R^d)$.
To check the two conditions \eqref{eq:tight-real} and \eqref{eq:tight-fourier}, let
$B_R(0)=\{x\in\R^d: |x|\le 1\}$ be the closed ball of radius $R$ and for a self-adjoint (vector-)operator $A$ let $\chi_R(A)=\id_{B_{R}(0)}(A)$ be
the associated orthogonal spectral projection and $\ol{\chi_R}(A):= \id - \chi_R(A)$ the
orthogonal projection onto the orthogonal complement of $\ran(\chi_R(A))$.
Then, for any $g\in L^2(\R^d)$,
 \bdm
    \int_{|x|>R} |g(x)|^2 \, dx  = \| \ol{\chi_R}(X) g\|_2^2
 \edm
where $X$ is the position operator, i.e., multiplication by $x$.
Since $\ol{\chi_R}(X)$ is an orthogonal projection in $L^2(\R^d)$, the triangle
inequality yields
 \bdm
    \| \ol{\chi_R}(X) f_n\| \le \|\ol{\chi_R}(X) f\| + \| \ol{\chi_R}(X) (f_n-f)\|
    \le \| \ol{\chi_R}(X) f\| + \| f_n-f\|
 \edm
So, since $f_n$ converges in norm to $f$,
 \bdm
    \limsup_{n\to\infty} \| \ol{\chi_R}(X) f_n\| \le \| \ol{\chi_R}(X) f\|
 \edm
for all $R>0$ and \eqref{eq:tight-real} follows by taking
the limit $R\to\infty$, since $\ol{\chi_R}(X)$ converges strongly to zero as $R\to\infty$.
\eqref{eq:tight-fourier} follows by an identical argument, using
 \bdm
    \int_{|k|>L} |\hatt{g}(k)|^2 \, dk  = \| \ol{\chi_L}(P) g\|^2
 \edm
with $P=-i\nabla$ the momentum operator in $L^2(\R^d)$. As above, we see
 \bdm
    \limsup_{n\to\infty} \| \ol{\chi_L}(P) f_n\| \le \| \ol{\chi_L}(P) f\|
 \edm
which implies \eqref{eq:tight-fourier} in the limit $L\to\infty$.

For the converse assume that $f_n$ converges weakly to $f\in L^2(\R^d)$ and that
\eqref{eq:tight-real} and \eqref{eq:tight-fourier} hold.
A judicious use of the triangle inequality reveals
 \bdm
    \begin{split}
    \|& f  -f_n\|
    \le \|\chi_R(X)(f-f_n)\| + \| \ol{\chi_R}(X)(f-f_n)\| \\
    &\le \|\chi_R(X) \chi_L(P)(f-f_n)\|
        + \|\chi_R(X) \ol{\chi_L}(P)(f-f_n)\| + \| \ol{\chi_R}(X)(f-f_n)\|\\
    &\le \|\chi_R(X) \chi_L(P)(f-f_n)\|
        + \|\ol{\chi_L}(P)(f-f_n)\| + \| \ol{\chi_R}(X)(f-f_n)\|\\
    &\le \|\chi_R(X) \chi_L(P)(f-f_n)\|
        +  \|\ol{\chi_L}(P) f\| + \| \ol{\chi_R}(X)f\|
        +  \|\ol{\chi_L}(P)f_n\| + \| \ol{\chi_R}(X)f_n\|
    \end{split}
 \edm
 Note that $\chi_R(X) \chi_L(P)$ is a Hilbert-Schmidt operator, in particular,
 a compact operator, see, for example, \cite{Birman-Solomjak-book,Simon-trace-ideals}.
 Thus it maps weakly convergent sequences into strongly convergent sequences. Hence
 $\lim_{n\to\infty}\|\chi_R(X) \chi_L(P)(f-f_n)\|=0$ since $f_n$ converges weakly to
 $f$. Taking the limit $n\to\infty$ in the above inequality one sees
 \bdm
    \limsup_{n\to\infty} \|f-f_n\|
    \le  \|\ol{\chi_L}(P) f\| + \| \ol{\chi_R}(X)f\|
        +  \limsup_{n\to\infty} \|\ol{\chi_L}(P)f_n\|
        + \limsup_{n\to\infty} \| \ol{\chi_R}(X)f_n\|
 \edm
 and then taking the limit $L,R\to\infty$ using $f\in L^2(\R^d)$ together with \eqref{eq:tight-real}
 and \eqref{eq:tight-fourier}, we get
 \bdm
    \limsup_{n\to\infty} \|f-f_n\|=0,
 \edm
 that is, $f_n$ converges strongly to $f$.
\epf
\begin{remark} That $\chi_R(X) \chi_L(P)$ is Hilbert-Schmidt is easy to see: Using the
    Fourier transform, one sees that $\chi_R(X) \chi_L(P)$ has an integral kernel with
    \beq\label{eq:int-kernel}
        \chi_R(X) \chi_L(P)(x,y)
        = \frac{1}{(2\pi)^{d/2}} \chi_R(x) \hatt{\chi_L}(x-y).
    \eeq
    Thus the Hilbert-Schmidt norm of $\chi_R(X) \chi_L(P)$ is given by
    \bdm
        \|\chi_R(X) \chi_L(P)\|_{\text{HS}}^2
        := \iint_{\R^d\times \R^d} |\chi_R(X) \chi_L(P)(x,y)|^2\, dx dy
        = (2\pi)^{-d/4} \|\chi_R\|_{L^2}^2 \|\chi_L\|_{L^2}^2
    \edm
    which is finite for all $0<R,L<\infty$.
\end{remark}

The discrete version of the strong convergence result is
\begin{lem}\label{lem:strong-convergence-lp} Let $1\le p<\infty$.
A sequence $(f_n)_{n\in\N}\subset l^p(\Z^d)$ is strongly converging to $f$ in $l^p(\Z^d)$
if and only if it is weakly convergent to $f$ and the sequence is tight, i.e.,
 \beq\label{eq:tight-discr}
   \lim_{L\to\infty} \limsup_{n\to\infty} \sum_{|x|>L} |f_n(x)|^p = 0.
 \eeq
\end{lem}
\begin{proof} The proof uses the same ideas in the continuous case. Let $K_L$ denote
the operator of multiplication with the characteristic function of the set
$\{x\in \Z^d:\, |x|\le L\}$,  that is,
 \bdm
   K_L f(x)= \left\{
                \begin{array}{cl}
                    f(x) & \text{ for } |x|\le L \\
                    0    & \text{ for } |x|> L
                \end{array}
             \right.
 \edm
and $\ol{K}_L:= \id-K_L$. Note that for all $L\ge 1$ both $K_L$ and $\ol{K}_L$ are bounded
operators on $l^p(\Z^d)$ with operator norm one since
 \bdm
    \|f\|_p^p = \|K_L f\|_p^p + \|\ol{K}_L f\|_p^p \ge \big[\max\big(\|K_L f\|_p, \|\ol{K}_L f\|_p  \big)\big]^p
 \edm
for all $f\in l^p(\Z^d)$. Moreover, the tightness-condition \eqref{eq:tight-discr} is equivalent to
 \beq\label{eq:tight-K}
   \lim_{L\to\infty}\limsup_{n\to\infty}\|\ol{K}_L f_n\|_p = 0 .
 \eeq

Since, for $1\le p<\infty$, $\ol{K}_L$ converges strongly to zero as $L\to\infty$, the proof of Lemma \ref{lem:strong-convergence-lp} is then as for the continuous case. In fact, it is much simpler since $K_L$ is even a finite range operator and thus trivially maps weakly converging sequences into
strongly convergent sequences.
\end{proof}

\begin{remark} The proof above breaks down for $p=\infty$, since for an arbitrary $f\in l^\infty(\Z^d)$,
one does not have
 \bdm
    \lim_{L\to\infty} \|\ol{K}_L f\|_\infty = 0,
 \edm
in general.  However, for the closed subspace $l^\infty_0(\Z^d)\subset l^\infty(Z^d)$ consisting of bounded sequences
indexed by $\Z^d$ vanishing at infinity, $\lim_{x\to\infty} f(x)=0$ for any $f\in l^\infty_0(\Z^d)$, the above
proof immediately generalizes and yields the analogous compactness statement: $(f_n)_{n\in\N}$ converges strongly
in $l^\infty_0(\Z^d)$ if and only if it converges weakly and
 \bdm
    \lim_{L\to\infty} \limsup_{n\to\infty} \sup_{|x|>L} |f_n(x)| = 0.
 \edm
\end{remark}

\section{Multi-linear estimates.} \label{app:multi-linear}
\setcounter{thm}{0}
\setcounter{equation}{0}
In this section we gather some a-priori bounds for the multi-linear functional
$\calQ_\mu^\mathrm{c}$, respectively $\calQ_\mu^\mathrm{d}$, which are used in the proof of
Lemma \ref{lem:fat-tail-bound-cont}, respectively Lemma \ref{lem:fat-tail-bound-discr}.
In the continuous case, the multi-linear bounds are an extension of our results
in \cite{HuLee2009}, in the discrete case the multi-linear bounds from \cite{HuLee-diffms}
are much stronger than their corresponding continuous counterparts.

\subsection{Multi-linear estimates for $\mathbf{\calQ_\mu^\mathrm{c}}$} \label{sec:mult-linear-cont}
We extend, with some small simplifications, the proofs of the multi-linear estimates from \cite{HuLee2009} from the case $\psi=\id_{[0,1]}$ to the more general densities needed here.
Again, we will write $\|f\|$ for the $L^2$-norm of a function $f\in L^2(\R)$
in this section.  As soon as the four-linear functional
$L^2(\R)^4\ni (f_1,f_2,f_3,f_4)\mapsto\calQ_\mu^\mathrm{c}(f_1,f_2,f_3,f_4)$ is bounded, scaling
$f\mapsto f/\sqrt{\lambda}$, shows that $P_\lambda^\mathrm{c}$ defined in \eqref{eq:max-cont}
obeys
 \beq\label{eq:scaling}
    P_\lambda^\mathrm{c}= P_1^\mathrm{c} \lambda^2
 \eeq
for all $\lambda>0$. In particular,  with $\lambda=\|f\|^2$,
 \beq\label{eq:boundedness-diag-cont}
   \calQ_\mu^\mathrm{c}(f,f,f,f)\le P_{\lambda}^\mathrm{c}= P_1^\mathrm{c} \|f\|^4.
 \eeq
For the boundedness we note
\begin{lem}\label{lem:boundedness-cont} Assume that $\mu$ has a density $\psi\in L^2(\R)$. Then
$ P_1^\mathrm{c} \le 12^{-1/4}\|\psi\| $ and for any functions $f_j\in L^2(\R)$, j=1,2,3,4,
 \beq\label{eq:boundedness-gen-cont}
  |\calQ_\mu^\mathrm{c}(f_1,f_2,f_3,f_4)| \le P_1^\mathrm{c} \prod_{j=1}^4 \| f_j\|  .
  \eeq
  Moreover, if $0\le \psi \not\equiv 0$, then $P_1^\mathrm{c}>0$.
\end{lem}
\begin{proof} We sketch the proof for the convenience of the reader. Using the triangle and
generalized H\"older inequalities,
  \begin{align*}\label{eq:Hoelder}
  | \calQ_\mu^\mathrm{c}(f_1,f_2,f_3,f_4) |
  & \leq \iint\limits_{\R\times\R} \prod _{j=1}^4 |T_tf_j| \, dx \psi(t)dt
    \leq \prod _{j=1}^4 \bigl(\iint\limits_{\R\times\R} |T_t f_j|^4 dx |\psi(t)|dt \bigr)^{1/4} \\
  & = \prod _{j=1}^4 \bigl( \calQ_{|\psi|}^\mathrm{c} (f_j, f_j, f_j, f_j )\bigr)^{1/4} .
  \end{align*}
Thus it is enough to show that
$\calQ_{|\psi|}^\mathrm{c} (f, f, f, f)\le P_1^\mathrm{c}\|f\|^4$
for some finite constant $P_1^\mathrm{c}$ and all $f\in L^2(\R)$.
Using the Cauchy-Schwarz inequality, one gets
 \bdm
 \begin{split}
    \calQ_\mu^\mathrm{c}(f,f,f,f)
    = \iint\limits_{\R\times\R} |T_tf|^{3+1} \psi(t) dx dt
    \le
    \bigl(\iint\limits_{\R\times\R} |T_tf|^6 dxdt\bigr)^{1/2}
    \bigl(\iint\limits_{\R\times\R} |T_tf|^2 \psi(t)^2\, dxdt \bigr)^{1/2} .
 \end{split}
 \edm
The second factor is seen to be bounded by $\|\psi\|\|f\|$ doing the $x$-integration
first, using that $T_t$ is a unitary operator on $L^2(\R)$. The first factor is bounded by
the one-dimensional Strichartz inequality,
 \beq\label{eq:Strichartz}
  \int_{\R}\int_{\R} |T_tf(x)|^6 \, dx dt \le S_1^6 \|f\|^6,
 \eeq
which holds due to the dispersive properties of the free
Schr\"odinger equation, \cite{GV85,Strichartz,SulemSulem}.  The sharp
constant in \eqref{eq:Strichartz} is known, $S_1= 12^{-1/12}$, one
even knows $S_2$ in two space dimensions, see \cite{Foschi,HZ06}.
Thus
 \bdm
  \calQ_\mu^\mathrm{c}(f,f,f,f)
 \le
  S_1^3 \|\psi\|\, \|f\|^4 .
 \edm
Hence $P_1^\mathrm{c}\le S_1^3\|\psi\|= 12^{-1/4}\|\psi\|$, using the sharp value for the
Strichartz constant. This proves the upper bound on $P_1^\mathrm{c}$.

To see that $P_1^\mathrm{c}>0$, note that $P_1^\mathrm{c}=0$ would imply
$\int_{\R}\int_{\R} |T_tf(x)|^4 \psi(t)\, dx dt = 0 $ for all $f\in L^2(\R)$.
Since  $\psi$ is non-negative,  $T_tf(x)$ has to be zero for Lebesgue almost every
$x$ and $\psi(t)dt$ almost every $t$. Thus, by the unicity of $T_t$,
    \bdm
        0= \int\int |T_{t}f(x)|^2\, dx \psi(t)dt = \int \|T_tf\|^2\, \psi(t)dt
         =  \|f\|^2 \int \psi(t)dt.
    \edm
Since $\int \psi(t)dt>0$ this shows $\|f\|=0$. Hence $P_1^\mathrm{c}>0$.
\end{proof}

\begin{remark}
    For a more explicit lower bound on $P_1^\mathrm{c}$ one can use a chirped Gaussian
   test-function similar to \cite{Kunze04, ZGJT01}, see also \cite{HuLee2009}.
   If the initial condition $f$ is given by
    \beq\label{eq:chirped-Gaussian}
     f(x)= A_0 e^{-x^2/\sigma_0} \quad \text{ with }\re(\sigma_0)>0
    \eeq
   then $u(t,x)= A(t)e^{-x^2/\sigma(t)}$ solves the free Schr\"odinger equation if
   $\sigma(t)= \sigma_0+4it$ and $A(t)= A_0\sqrt{\sigma_0}/\sqrt{\sigma(t)}$. Hence
    \beq
     T_tf(x) = A(t) e^{-x^2/\sigma(t)},
    \eeq
   for initial conditions of the form \eqref{eq:chirped-Gaussian},
   see, e.g., \cite{ZGJT01}. Thus
    \beq
     \int_{\R}\int_{\R} |T_tf|^4\, dx \psi(t)dt
     = \sqrt{\frac{\pi}{4}} |A_0|^4 \frac{|\sigma_0|^2}{\sqrt{\re \sigma_0}}
       \int_{\R} \frac{1}{|\sigma(t)|}\, \psi(t)dt .
    \eeq
   Choosing $|A_0|^2= \sqrt{2\re(\sigma_0)/(|\sigma_0|^2\pi)}$ yields
   the normalization  $\|f\|=1$ and hence
    \beq
     P_1^\mathrm{c}
    \ge \sup_{\sigma_0\in\C}
     \frac{\sqrt{\re(\sigma_0)}}{\sqrt{\pi}}
     \int_{\R} \frac{1}{\sqrt{\re(\sigma_0)^2+(\im(\sigma_0)+4t)^2}}\,\psi(t) dt
     >0.
    \eeq
    if $\psi$ is non-negative and positive on a set of positive Lebesgue measure.
\end{remark}

The proof of the multi-linear estimates is based on the by now well-known bilinear
Strichartz estimate, see, for example, \cite{Bourgain,CKSTT01,KPV91,OT98},
  \beq\label{eq:bilinear-Strichartz}
  \| T_tf_1 T_tf_2\|_{L^2(\R\times\R, dtdx)} \lesssim
  \frac{1}{\sqrt{\dist (\supp \hatt{f_1}, \supp \hatt{f_2})}}
  \| f_1\|_{L^2(\R)} \| f_2\|_{L^2(\R)} .
  \eeq
going back to \cite{Bourgain}. For a simple explicit proof of \eqref{eq:bilinear-Strichartz},
see for example, \cite{KMZ05} or \cite{HuLee2009}.
Now assume that the supports of $\hatt{f}_l$ and $\hatt{f}_m$ have positive distance for some
$l,m\in\{1,2,3,4\}$. Since
    \bdm
        |\calQ_\mu^\mathrm{c}(f_1,f_2,f_3,f_4)|
        \le
        \iint\limits_{\R\times\R} \prod_{j=1}^4 |T_tf_j(x)| \, |\psi(t)| dx dt
    \edm
we can assume that $l=1$ and $m=2$ without loss of generality. Then, by Cauchy-Schwarz followed by
\eqref{eq:bilinear-Strichartz} and \eqref{eq:boundedness-gen-cont},
    \bdm
    \begin{split}
        |\calQ_\mu^\mathrm{c}(f_1,f_2,f_3,f_4) |
        &\le
        \| T_tf_1 T_tf_2\|_{L^2(\R\times\R, dtdx)}
        \Big(\calQ_{|\psi|^2}(f_3,f_3,f_4,f_4) \Big)^{1/2}  \\
        &\lesssim
        \frac{\|\psi\|_{L^4}}{\sqrt{\dist (\supp \hatt{f_1}, \supp \hatt{f_2})}}
        \prod_{j=1}^4 \|f_j\|.
    \end{split}
    \edm
This yields the first part of the following
\begin{lem}[Refined multi-linear estimates]\label{lem:4linear-refined-cont}
Let $f_j\in L^2(\R)$ for $j=1,2,3,4$.\\[0.2em]
\itemthm Let the measure $\mu$ have density $\psi\in L^4(\R,dt)$. If, for some  $i \not=j$,
the supports of the Fourier transforms $\hatt{f_i}$ and $\hatt{f_j}$ are separated, i.e.,
$s= \dist (\supp \hatt{f_i}, \supp \hatt{f_j})>0$,  then
 \begin{equation}\label{eq:4linear-refined-Fourier}
 | \calQ_\mu^\mathrm{c}(f_1, f_2, f_3, f_4) |
 \lesssim
 \f{1}{\sqrt{s}}
 \|f_1\| \|f_2\|\|f_3\| \|f_4\|.
 \end{equation}
 where the implicit constant depends only on $\|\psi\|_{L^4(\R,dt)}$. \\
\itemthm Let the measure $\mu$ have density $\psi\in L^4(\R, t^2dt)$. If, for some $i \not=j$,
the supports of $f_i$ and $f_j$ are separated, i.e.,
$s= \dist (\supp f_i, \supp f_j)>0$, then
 \begin{equation}\label{eq:4linear-refined-real}
 | \calQ_\mu^\mathrm{c}(f_1, f_2, f_3, f_4) |
 \lesssim
 \f{1}{\sqrt{s}}
 \|f_1\| \|f_2\|\|f_3\| \|f_4\|.
 \end{equation}
 where the implicit constant in the bound depends only on $\|\psi\|_{L^4(\R,t^2dt)}$.
\end{lem}

It remains to show the second part of Lemma \ref{lem:4linear-refined-cont}. In fact,
using a symmetry of $\calQ_\mu^\mathrm{c}$ under Fourier transform, the second
half of the Lemma is \emph{equivalent} to the first half.
In order to formulate this symmetry we need a little bit more notation:
Given $f\in L^2(\R)$ let $\check{f}$ be its inverse
Fourier transform, that is, if $f\in L^1(\R)\cap L^2(\R)$, say,
 \beq\label{eq:inverse-Fourier}
   \check{f}(k)= \frac{1}{\sqrt{2\pi}} \int_{\R} e^{ikx} f(x)\, dx .
 \eeq
and given $\psi\in L^2(\R)$ define $\wti{\psi}$ by
 \beq\label{eq:psi-tilde}
 \wti{\psi}(\tau) = \frac{\psi(-1/(4\tau))}{2|\tau|} .
 \eeq
A simple change of variables shows
 \beq
  \|\wti{\psi}\| = \|\psi\|  ,
 \eeq
i.e., the $L^2$ norm is conserved.
\begin{lem}(Duality)\label{lem:duality}
Given a measure $\mu$ with density $\psi\in L^2(\R)$ let the measure $\wti{\mu}$
have density $\wti{\psi}$, where $\wti{\psi}$ is defined in \eqref{eq:psi-tilde}. Then
 \beq\label{eq:duality}
   \calQ_\mu^\mathrm{c}(f_1,f_2,f_3,f_4)
   = \calQ_{\wti{\mu}}^\mathrm{c}(\check{f_1},\check{f_2},\check{f_3},\check{f_4})
 \eeq
for all $f_j\in L^2(\R)$, j=1,2,3,4.
\end{lem}
We postpone the proof of Lemma \ref{lem:duality} for the moment. Given the duality \eqref{eq:duality},
the equivalence of \eqref{eq:4linear-refined-real} and \eqref{eq:4linear-refined-Fourier} is easy.
For example, assume that the supports of $f_1$ and $f_2$ are separated by at least $s$, that is,
the supports of the Fourier transforms of $\check{f_1}$ and $\check{f_2}$ are separated by at least
$s$. Hence \eqref{eq:duality} in tandem with \eqref{eq:4linear-refined-Fourier} yields
 \bdm
 \begin{split}
    | \calQ_\mu^\mathrm{c}(f_1, f_2, f_3, f_4) |
    =
    | \calQ_{\wti{\mu}}^\mathrm{c} (\check{f_1},\check{f_2},\check{f_3},\check{f_4}) |
    \lesssim
    \frac{1}{\sqrt{s}} \prod_{j=1}^4 \|\check{f_j}\|
    =
    \frac{1}{\sqrt{s}} \prod_{j=1}^4 \|f_j\|
 \end{split}
 \edm
where the implicit constant in the inequality depends only on
$\|\wti{\psi}\|_{L^4(\R,dt)}$. Since
$ \|\wti{\psi}\|_{L^4(\R,dt)} = 4^{1/4} \|\psi\|_{L^4(\R,t^2dt)}$
this proves \eqref{eq:4linear-refined-real}.
It remains to give the
\begin{proof}[Proof of Lemma \ref{lem:duality}]
The duality is a simple consequence of the so-called quasi-conformal symmetry of the
free Schr\"odinger evolution, which is true in all space dimensions, but we need only the
one-dimensional case here. The solution $u(t,x)= T_t f(x)$ of $i\partial_t u = - \partial_x^2 u$
with initial condition $u(0,x)=f(x)$ is given by
 \begin{align}
    u(t,x) &=
        \frac{1}{\sqrt{4\pi i t}} \int_{\R} e^{i\frac{|x-y|^2}{4t}} f(y)\, dy \label{eq:rep-real} \\
        &=
        \frac{1}{\sqrt{2\pi}} \int_{\R} e^{ixk} e^{-itk^2} \hatt{f}(k)\, dk . \label{eq:rep-Fourier}
 \end{align}
Expanding the square in \eqref{eq:rep-real} gives
 \bdm
    u(t,x) = \frac{1}{\sqrt{2it}} e^{i\frac{x^2}{4t}} \frac{1}{\sqrt{2\pi}}
    \int_{\R} e^{-i\frac{x}{2t}} e^{i \frac{y^2}{4t}} f(y)\, dy
 \edm
and comparing this with \eqref{eq:rep-Fourier}, one sees
 \beq\label{eq:quasi-conformal}
    u(t,x) = \frac{1}{\sqrt{2it}} e^{i\frac{x^2}{4t}} \wti{u}(-\frac{1}{4t}, -\frac{x}{2t})
 \eeq
where $\wti{u}(\tau, y)= T_\tau \check{f}(y)$ is the solution of the free
Schr\"odinger equation with initial condition $\check{f}$. Hence, setting $u_j(t,x)= T_tf_j(x)$,
$\wti{u}_j(\tau, y)= T_\tau \check{f_j}(y)$ and using \eqref{eq:quasi-conformal} one sees
 \beq
 \begin{split}
 \calQ_\mu^\mathrm{c}(f_1,f_2,f_3,f_4) &=
    \int_{\R}\int_{\R} \ol{u_1(t,x)} u_2(t,x) \ol{u_3(t,x)} u_4(t,x)\, dx \psi(t)dt \\
    &=
    \int_{\R}\int_{\R} \ol{\wti{u}_1(\tau,y)} \wti{u}_2(\tau,y) \ol{\wti{u}_3(\tau,y)} \wti{u}_4(\tau,y)
    \, dy \frac{\psi(-1/(4\tau))}{2|\tau|}d\tau \\
    &=
    \calQ_{\wti{\mu}}^\mathrm{c} (\check{f_1},\check{f_2},\check{f_3},\check{f_4})
 \end{split}
 \eeq
where we used \eqref{eq:quasi-conformal}, did the change of variables $x= 2t y$, $dx= 2|t|dy$,
and then $t= -1/(4\tau)$, $dt= d\tau/(4\tau^2)$, and defined $\wti{\psi}$ by \eqref{eq:psi-tilde}.
This proves \eqref{eq:duality}.
\end{proof}

As a last tool, we  note that although $\calQ_\mu^\mathrm{c}$ is certainly not local, it is `nearly' local in
the following sense.
\begin{lem}[Quasi-locality]\label{lem:quasi-locality}
 Assume that the measure $\mu$ has density $\psi\in L^2(\R)$. If, for some $s>0$, $f_1$ has
 support outside the interval $[-3s, 3s]$ and for $j=2,3,4$ all the functions $f_j$ have
 support within $[-s,s]$, then
 \beq
    \calQ_\mu^\mathrm{c}(f_1,f_2,f_3,f_4) = 0
 \eeq
 The same is true if $\hatt{f_1}$ has support outside the interval $[-3s, 3s]$ and
 for $j=2,3,4$ all the functions $\hatt{f_j}$ have support within $[-s,s]$
\end{lem}
\begin{proof} As in \cite{HuLee2009}, one uses the explicit representation \eqref{eq:rep-real} for the free time evolution to see that
 \beq\label{eq:rep-calQ}
 \begin{split}
    & \calQ_\mu^\mathrm{c}(f_1,f_2,f_3,f_4) \\
    & =
     \int_{\R}
     \f{\psi(t) d t}{(4\pi t)^2} \int _{\R} d x \int _{\R^4}
     e^{\f{ix(y_1 -y_2+y_3 -y_4)}{2t}}
     e^{\f{-i(y_1^2-y_2^2+y_3^2-y_4^2)}{4t}}
     \ol{f_1(y_1)}f_2(y_2)\ol{f_3(y_3)}f_4(y_4)\, d y \\
    & = \f{1}{8 \pi ^2}\int_{\R}
     \f{\psi(t) d t}{|t|} \int _{\R} d z \int _{\R^4}
     e^{i(y_1 -y_2+y_3 -y_4)z}
     e^{\f{-i(y_1^2-y_2^2+y_3^2-y_4^2)}{4t}}
     \ol{f_1(y_1)}f_2(y_2)\ol{f_3(y_3)}f_4(y_4)\, d y \\
    & = \f{1}{4 \pi}\int_{\R}
     \f{\psi(t) d t}{|t|} \int _{\R^4}
     \delta(y_1-y_2+y_3-y_4)
     e^{\f{-i(y_1^2-y_2^2+y_3^2-y_4^2)}{4t}}
     \ol{f_1(y_1)}f_2(y_2)\ol{f_3(y_3)}f_4(y_4) \, d y
 \end{split}
 \eeq
 where we first made the change of variables $x=2tz$, $dx= 2|t|dz$ and then used
 $\tfrac{1}{2\pi} \int_{\R} e^{iz\eta}\, dz = \delta(\eta)$, as distributions.
 Since the $y$-integration in formula \eqref{eq:rep-calQ} is restricted to the 3
 dimensional subspace given by $0= y_1-y_2+y_3-y_4$ one has $\calQ_\mu^\mathrm{c}(f_1,f_2,f_3,f_4)=0$
  whenever $f_1(y_1)f_2(y_2)f_3(y_3)f_4(y_4)=0$ on this subspace.
  This proves the first assertion.

 For the Fourier-space version, i.e., under the conditions stated on the Fourier
 transforms of $f_j$, simply note that the duality Lemma \ref{lem:duality} says
 \bdm
    \calQ_\mu^\mathrm{c}(f_1,f_2,f_3,f_4) = \calQ_{\wti{\mu}}^\mathrm{c}(\check{f_1},\check{f_2},\check{f_3},\check{f_4})
 \edm
 Thus \eqref{eq:rep-calQ} applied to $\calQ_{\wti{\mu}}^\mathrm{c}$ shows that
 $\calQ_\mu^\mathrm{c}(f_1,f_2,f_3,f_4)=0$ whenever the product
 $\check{f_1}(k_1)\check{f_2}(k_2)\check{f_3}(k_3)\check{f_4}(k_4)=0$
 on the subspace given by $k_1-k_2+k_3-k_4=0$.
 Since $\check{f}(k)= \hatt{f}(-k)$, this proves the second assertion of
 the Lemma.
\end{proof}
\begin{remarks}
\itemthm The proof of Lemma \ref{lem:quasi-locality} can be interpreted as a
    non-resonance effect: If the four wave-packets $f_1,f_2, f_3,f_4$ are
    \emph{non-resonant} in the sense that
    $\supp(f_1)-\supp(f_2)+\supp(f_3)-\supp(f_4)=0$ or
    $\supp(\hat{f}_1)-\supp(\hat{f}_2)+\supp(\hat{f}_3)-\supp(\hat{f}_4)=0$ then
    $\calQ_\mu^\mathrm{c}(f_1,f_2,f_3,f_4)=0$. \\[0.2em]
\itemthm \label{rem:regularity}
    Although the quasi-locality is not needed for our existence proof of maximizers,
    it is needed in the proof of regularity of weak solution of the dispersion management
    equation \eqref{eq:GT-weak-zero-cont}. Given Lemmata \ref{lem:boundedness-cont},
    \ref{lem:4linear-refined-cont}, and \ref{lem:quasi-locality}, a straightforward adaptation of
    the proof in \cite{HuLee2009}, which in our notation, was given there for the special case
    $\psi=\id_{[0,1]}$, shows that all weak solutions of the Gabitov-Turitsyn equation for
    vanishing average dispersion \eqref{eq:GT-weak-zero-cont} are already Schwartz functions as soon as
    $\psi\in L^2(\R)\cap L^4(\R)\cap L^4(\R,t^2dt)$.
\end{remarks}

\subsection{Multi-linear estimates for $\mathbf{\calQ_\mu^\mathrm{d}}$} \label{sec:mult-linear-discr}
In this section we will use $\|f\|$ for the $l^2$-norm of a sequence $f\in l^2(\Z)$.
Recall that $P_\lambda^\mathrm{d}$ is defined in $\eqref{eq:max-discr}$. As in
the continuous case, a simple scaling argument shows
 \beq\label{eq:scaling-discr}
    P_\lambda^\mathrm{d}= P_1^\mathrm{d} \lambda^2
 \eeq
for all $\lambda>0$. Thus,  with $\lambda=\|f\|^2$,
 \beq\label{eq:boundedness-diag-discr}
   \calQ_\mu^\mathrm{d}(f,f,f,f)\le P_{\lambda}^\mathrm{d}= P_1^\mathrm{d} \|f\|^4.
 \eeq

\begin{lem}\label{lem:boundedness-discr} Let $\mu$ be a bounded measure.
For any $f_j\in l^2(\Z^d)$, $j=1,2,3,4$, we have
 \beq \label{eq:boundedness-discr}
   |\calQ_\mu^\mathrm{d}(f_1,f_2,f_3,f_4)|   \le \mu(\R) \prod_{j=1}^4\|f_j\|,
 \eeq
 in particular, $0<P_\lambda^\mathrm{d}\le \mu(\R) \lambda^2$ for any $\lambda>0$.
\end{lem}
\bpf
By the triangle and H\"older inequalities
    \bdm
        |\calQ_\mu^\mathrm{d}(f_1,f_2,f_3,f_4)| \le
        \int_\R \sum_{x\in\Z^d} \prod_{j=1}^4 |S_t f_j(x)| \, \mu(dt)
        \le
        \int_\R \prod_{j=1}^4 \|S_t f_j\|_{l^4} \, \mu(dt)
    \edm
On the other hand, on the scale of $l^p$-spaces the simple but strong inequality
$\|g\|_{l^q}\le\|g\|_{l^p} $ holds for all $1\le p \le q\le \infty$. So
$\|S_t f_j\|_{l^4} \le \|S_t f_j\|_{l^2} = \|f_j\|_{l^2} $ since $S_t= e^{it\Delta}$
is unitary on $l^2(\Z)$.
The proof of $0<P_\lambda^\mathrm{d}$ is similar to the continuous case.
\epf

The following refined multi-linear estimate for $\calQ_\mu^\mathrm{d}$ is from
\cite{HuLee-diffms} where also the multi-dimensional case is done.
Note that the refined multi-linear estimate for $\calQ_\mu^\mathrm{d}$ shows a stronger decay than their continuous counterparts and as Lemma \ref{lem:boundedness-discr} this decay holds under the weakest possible assumption on $\mu$.
\begin{lem}\label{lem:4linear-refined-discr} Let $\mu$ be a bounded measure with bounded support
and assume that $s=\dist(\supp(f_k), \supp(f_l))>0$ for some $j,k\in \{1,2,3,4\}$. Then
 \beq\label{eq:4linear-refined-discr}
    |\calQ_\mu^\mathrm{d}(f_1,f_2,f_3,f_4)|
    \lesssim
    |s|^{-\delta |s|}  \prod_{j=1}^4 \|f_j\|
 \eeq
 for any $0<\delta<1/2$, where the implicit constant depends only on $\delta$.
\end{lem}
\bpf[Sketch of proof of Lemma \ref{lem:4linear-refined-discr}:]
The proof of Lemma \ref{lem:4linear-refined-discr} rests on the strong bilinear bound
 \beq\label{eq:strong-bilinear}
   \sup_{t\in[-\tau,\tau]} \| (e^{it\Delta}f_1) (e^{it\Delta} f_2) \|
   \lesssim
   s^{-\delta s}  \|f_1\| \|f_2\| \, .
 \eeq
 for all $0<\delta<1/2$, with $e^{it\Delta}$ the free discrete one-dimensional Schr\"odinger
 evolution and $s= \dist(\supp(f_1), \supp(f_2))$.
 Once one has \eqref{eq:strong-bilinear} the bound \eqref{eq:4linear-refined-discr} follows as in the
 continuous case.

 The estimate \eqref{eq:strong-bilinear} itself follows from the bound
 \beq\label{eq:strong-kernel}
   \sup_{t\in[-\tau,\tau]}|\la x|e^{it\Delta}|y\ra|
   \le \min(1, e^{4\tau}\frac{(4\tau)^{|x-y|}}{|x-y|!}  )
 \eeq
 for the kernel of the free time evolution $e^{it\Delta}$, $x,y\in\Z$ and $ 0\le \tau<\infty$.
 Here $\la x|M|y\ra = \la \delta_x, M\delta_y\ra$ for an operator $M$ on $l^2(\Z)$, where $\delta_x$ is the Kronecker delta--function.
 The bound \eqref{eq:strong-kernel} shows that unlike to the continuous case, the kernel of the free
 discrete Schr\"odinger evolution has a \emph{strong point--wise decay} locally uniformly in $t$. This is
 due to the finite speed of propagation for the discrete Schr\"odinger equation, or, in other words,
 the Fourier spectrum of the lattice $\Z$ is the bounded interval $[-\pi,\pi]$.
 The easiest way to see \eqref{eq:strong-kernel} is to note that since the discrete
 one-dimensional Laplace is bounded with norm  $\|\Delta\|=4$, the free discrete
 Schr\"odinger evolution can be written as a norm-converging
 exponential series
 $
    e^{it\Delta} = \sum_{n=0}^\infty  \frac{(it)^n}{n!} \Delta^n
 $.
 Thus
    \bdm
        |\la x|e^{it\Delta}|y\ra|
        \le
        \sum_{n=0}^\infty  \frac{(|t|)^n}{n!} |\la x| \Delta^n |y\ra|
        =
        \sum_{n=|x-y|}^\infty
        \frac{(4|t|)^n}{n!}
        \le e^{4\tau}\frac{(4\tau)^{|x-y|}}{|x-y|!}
    \edm
 since $\la x| \Delta^n |y \ra =0 $ if $|x-y|>n$ and $|\la x| \Delta^n |y\ra|\le \|\Delta\|^n\le 4^n$.
 For more details and extensions to $l^2(\Z^d)$ with $d>1$, see \cite{HuLee-diffms}.
\epf

\section{Shifts, boosts, and Galilei transformations.} \label{sec:Galilei}
\setcounter{thm}{0}
\setcounter{equation}{0}

We will only discuss the one-dimensional case which is somewhat easier than Galilei
transformations on $L^2(\R^d)$ since we do not have to deal with rotations in one dimension.
The unitary operator implementing the shift $S_y\xi :L^2(\R)\to L^2(\R)$,
$(S_\xi f)(x)= f(x-\xi)$ is given by
 \beq
    S_\xi= e^{-i\xi P}
 \eeq
where $P= -i\partial_x$ is the momentum operator. Indeed, since $e^{-i\xi P}$ corresponds
to multiplication by $e^{-i\xi k}$ in Fourier space, we have
 \bdm
    (e^{-i\xi P}f)(x) = \frac{1}{\sqrt{2\pi}} \int_\R e^{i(x-\xi)k}\hatt{f}(k)\, dk
    = f(x-\xi).
 \edm
Boosts, i.e., shifts in momentum space are given by $e^{iv\cdot}:L^2(\R)\to L^2(\R)$, i.e.,
multiplication by $e^{ivx}$, since
 \beq
    \hatt{e^{iv\cdot}f}(k)= \frac{1}{\sqrt{2\pi}} \int_\R e^{-ix(k-v)}f(x)\, dx
    = \hatt{f}(k-v).
 \eeq
Finally, if $G$ is a bounded (measurable) function then $G(P)$ is defined by
 \bdm
    \hatt{G(P)f}(k)= G(k) \hatt{f}(k) .
 \edm

Of course, for any $\xi\in\R$ the operators $G(P)$ and $e^{-i\xi P}$ commute,
$G(P)e^{-i\xi P}= e^{-i\xi P}G(P)$.  Moreover, for any $v\in\R$ the commutation relation
 \beq
    G(P)e^{iv\cdot} = e^{iv\cdot} G(P+v)
 \eeq
holds. Indeed, Computing the Fourier transform yields
 \bdm
 \begin{split}
    (G(P) e^{iv\cdot} f)^{\!\hatt{~}} (k)
    &= G(k) \hatt{e^{iv\cdot} f}(k) = G(k) \hatt{f}(k-v) \\
    &= ( G(\cdot+ v) \hatt{f} ) (k-v) = \big( G(P + v) f \big)^{\!\hatt{~}} (k-v) \\
    &= \big(e^{iv\cdot}G(P + v) f \big)^{\!\hatt{~}} (k).
 \end{split}
 \edm
In particular, choosing $G(P)= e^{-itP^2}$, we see the commutation relation
 \beq\label{eq:commutation}
 \begin{split}
    e^{-itP^2} e^{iv\cdot} e^{-i\xi P}
    & = e^{iv\cdot} e^{-i\xi P} e^{-it(P+v)^2}
        = e^{iv\cdot} e^{-i\xi P} e^{-it(P^2+2vP + v^2)} \\
    &= e^{-itv^2} e^{iv\cdot} e^{-i(\xi +2tv)P} e^{-itP^2} .
 \end{split}
 \eeq
Now let $f\in L^2(\R)$. Then $u(t)= T_tf= e^{-itP^2}f$ is the solution of the
(one-dimensional) Schr\"odinger equation $-i\partial_t u = P^2 u = -\partial_x^2 u$
with initial condition $u(0)= f$. Using \eqref{eq:commutation}, the solution of the
free Schr\"odinger equation for the translated and boosted initial condition
$f_{\xi,v}= e^{iv\cdot}e^{-i\xi P} f$ is given by
 \beq\label{eq:Galilei-trafo}
 \begin{split}
    u_{\xi,v}(t,x)
    &:=  e^{-itP^2}f_{\xi,v} (t,x)
        = \big(e^{-itP^2} e^{iv\cdot}e^{-i\xi P} f\big)(x) \\
    &= \big(e^{-itv^2} e^{iv\cdot} e^{-i(\xi +2tv)P} e^{-itP^2} f\big)(x) \\
    &= e^{-itv^2} e^{iv x} \big(e^{-i(\xi +2tv)P} e^{-itP^2} f\big)(x) \\
    &   = e^{-itv^2} e^{iv x} \big( e^{-itP^2} f\big)(x - \xi -2tv) \\
    &= e^{-itv^2} e^{iv x} u(t,x - \xi -2tv),
 \end{split}
 \eeq
that is, on the level of the solutions of the free time-dependent Schr\"odinger equation,
translations and boost of the initial condition are implemented by the Galilei transformations $\calG_{\xi,v}$ given by
$(\calG_{\xi,v}u)(t,x) := u_{\xi,v}(t,x)= e^{-itv^2} e^{iv x} u(t,x - \xi -2tv)$.
Except for the time-dependent phase factor $e^{-itv^2}$, formula \eqref{eq:Galilei-trafo} is
exactly what one would have guessed from classical mechanics

Note that  $P^2= -\Delta$. A simple calculation now shows that the functional
    \bdm
        f\mapsto \calQ_\mu^\mathrm{c}(f,f,f,f)=\int_\R\int_\R |(e^{-itP^2}f)(x)|^4\, dx \mu(dt)
    \edm
is invariant under translations and boosts in $L^2(\R)$. Similarly,
it is straightforward to see that the 4-linear functional $f_j\mapsto \calQ_\mu^\mathrm{c}(f_1,f_2,f_3,f_4)$
is invariant under simultaneous shifts and boosts of the $f_j$.

\textbf{Acknowledgements: }
It is  a pleasure to thank Vadim Zharnitsky
for instructive discussions on the dispersion management technique.
Dirk Hundertmark thanks
the Max-Planck Institute for Physics of Complex Systems in Dresden
and the Max-Planck Institute for Mathematics in the Sciences in Leipzig for their warm hospitality while part of this work was done.
\renewcommand{\thesection}{\arabic{chapter}.\arabic{section}}
\renewcommand{\theequation}{\arabic{chapter}.\arabic{section}.\arabic{equation}}
\renewcommand{\thethm}{\arabic{chapter}.\arabic{section}.\arabic{thm}}

\def\cprime{$'$}


\begin{thebibliography}{100}
\small{

\bibitem{AB98} M.J.\ Ablowitz and G.\ Biondini,
    \textit{Multiscale pulse dynamics in communication systems
    with strong dispersion management.}
    Opt.\ Lett.\ \textbf{23} (1998), 1668--1670.

\bibitem{AM01} M.\ Ablowitz and Z.\ H.\ Musslimani,
    \textit{Discrete Diffraction Managed Spatial Solitons.}
    Phys.\ Rev.\ Lett.\  \textit{87} (2001), 254102 [4 pages].

\bibitem{AM02} M.\ Ablowitz and Z.\ H.\ Musslimani,
    \textit{Discrete vector spatial solitons in a nonlinear waveguide array.}
    Phys.\ Rev.\ E \textbf{65} (2002), 056618.

\bibitem{AM03} M.\ Ablowitz and Z.\ H.\ Musslimani,
    \textit{Discrete spatial solitons in a diffraction managed nonlinear
            waveguide array: a unified approach.}
    Physica D \textbf{184} (2003), 276–-303.


\bibitem{Birman-Solomjak-book}
    M.~Sh.~Birman and M.~Z.~Solomjak,
    \textit{Spectral theory of selfadjoint operators in Hilbert space.}
    Translated from the 1980 Russian original by S.~Khrushchëv and V.~Peller.
    Mathematics and its Applications (Soviet Series).
    D.~Reidel Publishing Co., Dordrecht, 1987. xv+301 pp.


\bibitem{Bourgain}
    J.~Bourgain, \textit{Global solutions of nonlinear Schr\"odinger equations.}
    American Mathematical Society Colloquium Publications, \textbf{46}.
    AMS, Providence, RI, 1999.


\bibitem{CBG-J94} D.\ Cai, A.\ R.\ Bishop, and N.\ Groenbech-Jensen,
    \textit{Localized states in discrete nonlinear Schr\"odinger equations.}
    Phys.\ Rev.\ Lett.\ \textbf{72} (1994), 591–-595.


\bibitem{Cazenave03} Thierry Cazenave,
    \textit{Semilinear Schr\"odinger equations,}
    Courant Lecture Notes in Mathematics \textbf{10},
    AMS, Providence, Rhode Island, 2003.


\bibitem{CaLi82} T.~Cazenave and   P.-L.~Lions,
    \textit{Orbital stability of standing waves for some nonlinear Schr\"odinger equations,}  Comm.\ Math.\ Phys.\ \textbf{ 85} (1982), no.\ 4, 549--561.


\bibitem{CGTD93}  A.~R.~Chraplyvy, A.~H.~Gnauck,  R.~W.~Tkach, R.~M.~Derosier,
    \textit{8~10 Gb/s transmission through 280 km of dispersion-managed fiber.}
    IEEE Phot.\ Tech.\ Lett.\ \textbf{5} (1993), 1233--1235.


\bibitem{Chraplyvy-etal95}
    A.~R.~Chraplyvy,   A.~H.~Gnauck,  R.~W.~Tkach, R.~M.~Derosier, E.~R.~Giles, B.~M.~Nyman,
    G.~A.~Ferguson, J.~W.~Sulhoff, J.~L.~Zyskind,
    \textit{One-third terabit/s transmission through 150 km of dispersion-managed fiber,}
    Photonics Technology Letters, IEEE (1995), \textbf{7},  Issue: 1, 98-100.


\bibitem{CJ88} D.~N.\ Christodoulides and R.~I. Joseph,
    \textit{ Discrete self-focusing in nonlinear arrays of coupled waveguides.}
    Opt.\ Lett.\ \textbf{13} (1988), 794–-796.


\bibitem{CKSTT01} J.\ Colliander, M.\ Keel, G.\ Staffilani, H.\ Takaoka, and T.\ Tao,
    \textit{A refined global well-posedness result for Schr\"odinger equations
        with derivative.}
        SIAM J.\ Math.\ Anal.\ \textbf{34} (2002), 64--86.


\bibitem{Davydov73} A.\ S.\ Davydov,
    \textit{Theory of contraction of proteins under their excitation.}
    J.\ Theor.\ Biol.\ \textbf{38} (1973), 559–-569.


\bibitem{ESMBA98} H.\ Eisenberg, Y.\ Silverberg, R.\ Morandotti, A.\ Boyd, and J.\ Aitchison,
    J. (1998). \textit{Discrete spatial optical solitons in waveguide arrays.}
    Phys.\ Rev.\ Lett.\ \textit{81} (1998), 3383-–3386.


\bibitem{ESMA00} H.\ Eisenberg, Y.\ Silverberg, R.\ Morandotti, and J.\ Aitchison,
    \textit{Diffraction management.}
    Phys. Rev. Lett. \textbf{85} (2000), 1863–-1866.


\bibitem{Ekeland72} I.\ Ekeland,
    \textit{Sur les probl\`emes variationnels.} (French)
    C.\ R.\ Acad.\ Sci.\ Paris S\'er.\ A-B \textbf{275} (1972), A1057--A1059.

\bibitem{Ekeland73} I.\ Ekeland,
    \textit{Remarques sur les probl\`emes variationnels.} (French)
    C.\ R.\ Acad.\ Sci.\ Paris S\'er. A-B  \textbf{276}  (1973), A1347--A1348.


\bibitem{EHL2009} M.~B.~Erdo\smash{\u g}an, D.~Hundertmark, and Y.-R.~Lee,
    \textit{Exponential decay of dispersion managed solitons for vanishing average dispersion,}
    preprint.


\bibitem{Foschi} Damiano Foschi,
    \textit{Maximizers for the Strichartz inequality.}
    J.\ Eur.\ Math.\ Soc.\ \textbf{9} (2007), 739--774.


\bibitem{GT96a} I.\ Gabitov and S.K.\ Turitsyn,
    \textit{Averaged pulse dynamics in a cascaded transmission system
    with passive dispersion compensation.}
    Opt.\ Lett.\  \textbf{21} (1996), 327--329.

\bibitem{GT96b} I.\ Gabitov and S.K.\ Turitsyn,
    \textit{Breathing solitons in optical fiber links.}
    JETP Lett.\ \textbf{63} (1996) 861.


\bibitem{GV85} Jean~Ginibre and Giorgio~Velo,
    \textit{The global Cauchy problem for the nonlinear Schr\"odinger equation.}
    Ann.\ Inst.\  H.\ Poincar\'{e} Anal.\ Non Lin\'{e}aire
    \textbf{2} (1985), 3009--327.

\bibitem{HuLee2009} D.\ Hundertmark, and Y.-R.\ Lee,
    \textit{Decay estimates and smoothness for solutions of the dispersion managed
    non-linear Schr\"odinger equation.}
    Comm.\ Math.\ Phys.\ \textbf{286} (2009), no.\ 3, 851--873.


\bibitem{HuLee-diffms} D.\ Hundertmark, and Y.-R.\ Lee,
    \textit{Super-exponential decay of diffraction management solitons,}
    preprint.

\bibitem{HZ06} Dirk Hundertmark and Vadim Zharnitsky,
    \textit{On sharp Strichartz inequalities for low dimensions.}
    International Mathematics Research Notices, vol.\ 2006, Article ID 34080, 18 pages, 2006.
    doi:10.1155/IMRN/2006/34080


\bibitem{Jabri} Y.\ Jabri,
    \textit{The mountain pass theorem. Variants, generalizations, and some applications.}
    Encyclopedia of Mathematics and its Applications \textbf{95}.


\bibitem{KPV91} C.E.\ Kenig, G.\ Ponce, and L.\ Vega,
    \textit{Oscillatory integrals and regularity of dispersive
    equations.}
    Indiana Univ.\ Math.\ J.\ \textbf{40} (1991), 33--68.


\bibitem{KH97} S.\ Kumar and A.\ Hasegawa,
    \textit{Quasi-soliton propagation in dispersion-managed optical fibers.}
     Opt.\ Lett.\ \textbf{ 22} (1997), 372--374.



\bibitem{Kunze04} Markus Kunze,
    \textit{On a variational problem with lack of compactness related to
    the Strichartz inequality.}
    Calc.\ Var.\ Partial Differential Equations\
    \textbf{19}  (2004),  no.\ 3, 307--336.

\bibitem{KMZ05}
    M.~Kunze, J.~Moeser, and V.~Zharnitsky,
    \textit{Ground states for the higher-order dispersion managed NLS equation in
    the absence of average dispersion,}
    J.\ Differential Equations \textbf{209} (2005), no.\ 1, 77--100.


\bibitem{Kurtzge93} Kurtzke,
    \textit{Suppression of fiber nonlinearities by appropriate
    dispersion management.} IEEE Phot.\ Tech.\ Lett.\ \textbf{5} (1993), 1250--1253.


\bibitem{LK98} T.\ Lakoba and D.J.\ Kaup,
    \textit{Shape of the stationary pulse in the strong
    dispersion management regime.}
    Electron.\ Lett.\ \textbf{34} (1998), 1124--1125.


\bibitem{LandauLifshitz}
    L.D.\ Landau and E.M.\ Lifshitz,
    \textit{Course of theoretical physics. Vol. 1. Mechanics.}
    Third edition. Pergamon Press, Oxford-New York-Toronto, Ont., 1976.


\bibitem{LiebLoss}
    E.~H.~Lieb and M.~Loss, \textit{Analysis.} Second edition.
    Graduate Studies in Mathematics, \textbf{14}.
    AMS, Providence, RI, 2001.


\bibitem{LKC80} C.\ Lin, H.\ Kogelnik, and L.\ G.\ Cohen,
    \textit{Optical pulse equalization and low dispersion transmission in singlemode
            fibers in the 1.3--1.7 $\mu$m spectral region.}
    Opt.\ Lett.\ \textbf{5} (1980), 476–-478.



\bibitem{Lions84} P.~L.\ Lions,
    \textit{The concentration-compactness principle in the calculus of
    variations. The locally compact case, part 1 and 2.}
    Annales de l'institut Henri Poincar\'e (C) Analyse non lin\'eaire\
    \textbf{1} no.\ 2 and no.\ 4 (1984), 109--145 and 223--283


\bibitem{Lushnikov04} Pavel M.\ Lushnikov,
    \textit{Oscillating tails of dispersion-managed soliton.}
    J.\ Opt.\ Soc.\ Am.\ B\ \textbf{21} (2004), 1913--1918.


\bibitem{MM99} P.V.\ Mamyshev and N.A.\ Mamysheva,
    \textit{Pulseoverlapped dispersion-managed data transmission
    and intrachannel four-wave mixing.}
    Opt.\ Lett.\ \textbf{24} (1999) 1454--1456.


\bibitem{Moeser05} Jamison Moeser,
    \textit{Diffraction managed solitons: asymptotic validity and excitation thresholds}
    Nonlinearity \textbf{18} (2005), 2275–-2297.



\bibitem{MGLWXK03} L.F.\ Mollenauer, A.\ Grant, X.\ Liu, X.\ Wei, C.\ Xie, and I.\ Kang,
    \textit{Experimental test of dense wavelengthdivision multiplexing using novel,
    periodic-group-delaycomplemented dispersion compensation and
    dispersionmanaged solitons.}
    Opt.\ Lett.\ \textbf{28} (2003), 2043--2045.


\bibitem{MMGNMG-NV99} L.F. Mollenauer, P.V. Mamyshev, J. Gripp, M.J. Neubelt, N.
    Mamysheva, L. Gr\"uner-Nielsen, and T. Veng,
    \textit{Demonstration of massive wavelength-division multiplexing over
    transoceanic distances by use of dispersionmanaged solitons.}
    Opt.\ Lett.\ \textbf{ 25} (1999), 704--706.


\bibitem{OT98} T.~Ozawa and Y.~Tsutsumi,
    \textit{Space-time estimates for null gauge forms and nonlinear
    Schr\"odinger equations.}
    Differential Integral Equations\ \textbf{11} (1998), 201--222.



\bibitem{Panayotaros05} P.\ Panayotaros,
    \textit{Breather solutions in the diffraction managed NLS equation.}
    Physica D \textbf{206} (2005), 213–-231.


\bibitem{ReSi4}
    M.~Reed and B.~Simon,
    \textit{Methods of modern mathematical physics. IV. Analysis of operators.}
    Academic Press, New York--London, 1978.


\bibitem{Scott85} A.\ C.\ Scott, 1985 \textit{Davydov solitons in polypeptides}
    Phil.\ Trans.\ R.\ Soc.\ Lond.\ \textit{A 315} (1985), 423–-436.



\bibitem{Simon-trace-ideals}
    B.~Simon, \textit{Trace ideals and their applications.} Second edition.
    Mathematical Surveys and Monographs, \textbf{120}.
    AMS, Providence, RI, 2005. viii+150 pp.


\bibitem{Stanislavova05} Milena Stanislavova,
    \textit{Regularity of ground state solutions of DMNLS equations.}
    J.\ Diff.\ Eq.\ \textbf{210} (2005), no.\ 1, 87--105.

\bibitem{Stanislavova07} Milena Stanislavova,
    \textit{Diffraction Managed Solitons with Zero Mean Diffraction}
    Journal of Dynamics and Differential Equations \textbf{ 19} (2007),
    no.\ 2., 295--307.


\bibitem{Strichartz}
    R.~S.~Strichartz,
    \textit{Restrictions of Fourier transforms to quadratic surfaces and
        decay of solutions of wave equations,}
    Duke Math.\ J.\ \textbf{44} (1977), 705--714.


\bibitem{Strichartz00} R.\ S.\ Strichartz,
    \textit{The Way of Analysis.} Revised Edition.
    Jones and Bartlett publications, (2000).



\bibitem{SKES03} A.\ Sukhorukov, Y.\ Kivshar, E.\ H.\ Eisenberg, and Y.\ Silberberg,
    \textit{Spatial optical solitons in waveguide arrays.}
    IEEE J.\ Quantum Electron.\ \textit{39} (2003), 31–-50.

\bibitem{SulemSulem}
    Cathrine Sulem and Pierre-Louis Sulem,
    \textit{The non-linear Schr\"odinger equation. Self-focusing and wave collapse.}
    Applied Mathematical Sciences, \textbf{139}. Springer-Verlag, New York, 1999.


\bibitem{TS98} A.\ Trombettoni and A.\ Smerzi,
    \textit{Discrete solitons and breathers with dilute Bose\-–Einstein condensates.}
    Phys.\ Rev.\ Lett.\ \textit{86} (1998), 2353–-2356.


\bibitem{TDNMSF99}
 S.K.\ Turitsyn, N.J.\ Doran, J.H.B.\ Nijhof, V.K.\ Mezentsev,
 T.\ Sch\"afer, and W.\ Forysiak,
 in \textit{Optical Solitons: Theoretical challenges and industrial perspectives, }
 V.E.\ Zakharov and S.\ Wabnitz, eds.\ (Springer Verlag, Berlin, 1999), p.\ 91.

\bibitem{Tetal03} S.\ K.\ Turitsyn, E.\ G.\ Shapiro, S.\ B.\ Medvedev, M.\ P.\ Fedoruk,
    and V.\ K.\ Mezentsev,
    \textit{Physics and mathematics of dispersion-managed optical solitons,}
    Comptes Rendus Physique, Acad\'emie des sciences/\'Editions scientifiques
    et m\'edicales \textbf{4} (2003), 145--161.

\bibitem{Weinstein-86} M.~I.~Weinstein,
    \textit{Lyapunov stability of ground states of nonlinear dispersive
    evolution equations,}
    Comm.\ Pure Appl.\ Math.\  \textbf{39}  (1986),  no.\ 1, 51--67.

\bibitem{Weinstein99} M.~I.~Weinstein,
    \textit{Excitation thresholds for nonlinear localized modes on lattices,}
    Nonlinearity \textbf{12} (1999), 673-–91.



\bibitem{ZGJT01}
 V.\ Zharnitsky, E.\ Grenier, C.K.R.T. Jones, and S.K.\ Turitsyn,
 \textit{Stabilzing effects of dispersion management,}
 Physica D \textbf{152-153} (2001), 794--817.

}
\end{thebibliography}
\end{document}